\documentclass{aastex}
\usepackage{emulateapj5}
\usepackage{bm}

\newcommand {\apgt} {\ {\raise-.5ex\hbox{$\buildrel>\over\sim$}}\ }
\newcommand {\aplt} {\ {\raise-.5ex\hbox{$\buildrel<\over\sim$}}\ }

\shorttitle{The formation and evolution of young star clusters}
\shortauthors{M. S. Fujii \& S. Portegies Zwart}

\begin{document}


\title{The formation and dynamical evolution of young star clusters}


\author{M. S. Fujii\altaffilmark{1}}
\affil{Division of Theoretical Astronomy, 
   National Astronomical Observatory of Japan, 2--21--1 Osawa,
   Mitaka, Tokyo 181--8588}

\author{S. Portegies Zwart\altaffilmark{2}}
\affil{Leiden Observatory, Leiden University, PO Box 9513, 2300 RA,
  Leiden, The Netherlands}
\email{michiko.fujii@nao.ac.jp}


\altaffiltext{1}{NAOJ Fellow}


\begin{abstract}
Recent observations have revealed a variety of young star clusters,
including embedded systems, young massive clusters, and
associations. We study the formation and dynamical evolution of 
these clusters using a combination of simulations and theoretical models.  
Our
simulations start with a turbulent molecular cloud that collapses
under its own gravity. The stars are assumed to form in the densest 
regions in the collapsing cloud after an initial free-fall times of 
the molecular cloud. The
dynamical evolution of these stellar distributions are continued by
means of direct $N$-body simulations.  
The molecular clouds typical for the Milky Way
Galaxy tend to form embedded clusters which evolve to resemble open
clusters. The associations were initially considerably more clumpy, 
but lost their irregularity in about a dynamical time scale
due to the relaxation process. The densest molecular
clouds, which are absent in the Milky Way but are typical in
starburst galaxies, form massive young star clusters. They
indeed are rare in the Milky Way.
Our models indicate a distinct evolutionary path from molecular clouds
to open clusters and associations or to massive
star clusters.  The mass-radius relation for both types of evolutionary
tracks excellently matches the observations.  According to our
calculations the time evolution of the half-mass radius for open
clusters and associations follows $r_{\rm h}/{\rm pc}=2.7(t_{\rm age}/{\rm
  pc})^{2/3}$, whereas for massive star clusters $r_{\rm h}/{\rm
  pc}=0.34(t_{\rm age}/{\rm Myr})^{2/3}$. Both trends are consistent with the
observed age-mass-radius relation for clusters in the Milky Way.
\end{abstract}


\keywords{galaxies: star clusters: general  --- (Galaxy:) open clusters and associations: general --- methods: numerical}

\section{Introduction}

Star clusters are classically categorized in two groups, Galactic open
clusters and globular clusters. Open clusters are generally rather
young ($\lesssim 1$\,Gyr) with typically $100-10^4$ stars, hereafter we
call them ``classical'' open clusters. Globular clusters are old ($\gtrsim
10$ Gyr), more massive ($\gtrsim 10^5M_{\odot}$), and dense ($\gtrsim
100M_{\odot}{\rm pc}^{-3}$).  Recent observations indicate that there
is a wide variety among open star clusters in the Milky Way. These
types include
\begin{itemize}
\item embedded clusters, which are very young $\lesssim3$ Myr and
  therefore still embedded in their natal gas cloud
  \citep{2003ARA&A..41...57L}.  Embedded clusters reside in the
  Galactic disk and are composed of several 100 stars in a volume with a
  radius of $\sim 1$ pc \citep{2002AJ....124.2739F}.
\item associations, which are considered unbound from the
moment they were born \citep{2011MNRAS.410L...6G}.
\item massive clusters, which are also young
($\lesssim 10$ Myr) and extremely dense ($\gtrsim 10^3M_{\odot}{\rm
  pc}^{-3}$) \citep{2010ARA&A..48..431P}.
\end{itemize}
Some of embedded clusters evolve into classical open clusters, if 
they survive gas expulsion \citep{2003ARA&A..41...57L,2015PASJ...67...59F}.  

Young massive clusters are common in nearby starburst galaxies such as
in M83 \citep{2011MNRAS.417L...6B} and M51
\citep{2011ApJ...727...88C}, but they are rare in the Milky Way. Two
massive young star clusters reside close to the Galactic center, i.e.,
Arches and Quintuplet, and the others are in the spiral arms. This
latter category includes the clusters NGC 3603, Westerlund 1 and 2, and
Trumpler 14 \citep{2010ARA&A..48..431P}.

\citet{2009A&A...498L..37P} suggested another type of young star
clusters, ``leaky clusters.''  Leaky clusters have a mass similar to
those of the massive clusters ($\sim 10^4M_{\odot}$), but with a much
lower density ($\sim $1--10$M_{\odot}{\rm pc}^{-3}$).
\citet{2010ARA&A..48..431P} classified the leaky clusters listed in
\citet{2009A&A...498L..37P} as OB associations.

According to the argumentation in \citet{2011MNRAS.410L...6G} the
distinction between an open cluster and an association can be made on
the ratio between the age of the stars and dynamical time of the
system ($t_{\rm age}/t_{\rm dyn}$). If the age of the stars exceed the
dynamical age of the system, the stars must be bound
together. Otherwise the system is unbound.

In an attempt to clarify the various classes and families of stellar
conglomerates we discuss, in this paper, the formation and dynamical
evolution of young star clusters by means of simulations. The
numerical modeling used here allows us to make a more clear
distinction between the difference in initial conditions and the
difference in evolution. It therefore helps us to differentiate
between the various classes and families of clustered stellar
environments.

In previous papers, we performed direct $N$-body simulations using
initial conditions constructed from the results of hydrodynamical
simulations of turbulent molecular clouds. There we found that young
massive clusters form from turbulent molecular clouds, if the local
star formation efficiency depends on the local gas density
\citep{2015MNRAS.449..726F,2015PASJ..tmp..163F}.  We also found that
observed embedded clusters tend to evolve into classical open clusters
\citep{2015PASJ..tmp..163F}. Our simulations, however, did not provide 
a channel for forming associations (or leaky clusters, according to
\cite{2009A&A...498L..37P}).

At this point it is still unclear how leaky clusters form.
\citet{2011A&A...536A..90P} proposed that leaky clusters are born as
embedded clusters, that their mass increases due to a prolonged
phase of star formation, and that the expansion is driven by the
expulsion of the residual gas.  This scenario was tested by means of
simulations in \citet{2013A&A...559A..38P},
\citet{2013A&A...549A.132P}, and \citet{2014ApJ...794..147P}, in which
it was concluded that the known embedded clusters in the Galactic disk
are the ancestors of leaky clusters.

In our previous simulations we did not find leaky clusters.  This may
have been a result of our selected initial conditions for the parental
molecular cloud, for which we chose rather massive
($10^5$--$10^6\,M_{\odot}$) and dense ($100$--$1000\,{\rm cm}^{-3}$)
structures.  The molecular clouds observed in the Milky Way tend to
follow Larson's relation \citep{1981MNRAS.194..809L}, which indicates a
relation between cloud mass and density: According to this law massive
clouds have a lower density, if the clouds are close to be virialized.  
The initial conditions in our previous
study would then biased towards too dense clouds compared to the
typical massive clouds in the Milky Way.

In this paper, we expand on the initial parameter space, by also
allowing massive clouds with a lower density.  This expansion of the
parameter space helps in the formation of associations, as well as for
making dense massive clusters.  We support our numerical models with
theoretical arguments in order to understand the the dynamical
evolution of each type of star clusters (classical open, embedded,
young massive, and leaky clusters or associations).

\section{Simulations}

We perform a series of $N$-body simulations based on the results of
hydrodynamical simulations of turbulent molecular clouds.  We first
perform simulations of molecular clouds with a turbulent velocity
field using an smoothed particle hydrodynamics (SPH) code.  The
resolution of the hydrodynamical simulations is relatively low and
therefore the simulation cannot resolve the formation of individual
stars, but can resolve the clumpy structures of the gas.  After around
one free-fall time of the initial molecular clouds, we stop the
hydrodynamical simulations and replace a part of gas particles with
stellar particles assuming a star formation efficiency depending on
the local density.  We then remove all residual gas particles and
perform direct $N$-body simulations only with stellar particles. We
describe the details of the initial conditions and the simulations in
the following \citep[see
  also][]{2015MNRAS.449..726F,2015PASJ..tmp..163F}.

\subsection{The Astronomical Multipurpose Software Environment}
The hydrodynamical simulations and the data analyses 
in this study are performed using the AMUSE
framework \citep{2013CoPhC.183..456P,2013A&A...557A..84P}.  AMUSE is
not a single code, but a extensive library of more than 50
high-performance simulation codes.  The AMUSE consortium is a spin-off
from the MODEST community, which upon three workshop in Lund,
Amsterdam, and Split culminated in a first implementation of, what at
that time was called the Multi-User Software Environment (or MUSE)
\citep{2009NewA...14..369P}. Later the package was extended from its
primary objective of Noah's Arc (two codes per domain) to about a
dozen codes per domain.

Apart from scientific production software, AMUSE also supports from
generating initial conditions to data processing. The fundamental
package is written in the Python language and it is freely available
via {\tt Github} and via the project web page at
\url{http://amusecode.org}.  All the scripts used to run the
simulations in this paper are available via this project web page.

\subsection{Hydrodynamical Simulations}

\subsubsection{Initial Conditions for Molecular Clouds}
All initial conditions are generated using the AMUSE framework.
We adopt isothermal (30K) homogeneous spheres as initial conditions of
molecular clouds following \citet{2003MNRAS.343..413B}.  We give a
divergence-free random Gaussian velocity field $\delta \bm{v}$ with a
power spectrum $|\delta v|^2\propto k^{-4}$
\citep{2001ApJ...546..980O,2003MNRAS.343..413B}.  The spectral index
of $-4$ appears in the case of compressive turbulence (Burgers
turbulence), and recent observations of molecular clouds
\citep{2004ApJ...615L..45H} and numerical simulations
\citep{2010A&A...512A..81F, 2011ApJ...740..120R,2013MNRAS.436.1245F}
also suggested values similar to $-4$. Each model is run with a
different random seed for a realization of the initial
conditions.

We adopt the virial ratio $|E_{\rm k}|/|E_{\rm p}| = 1$ (here $E_{\rm
  k}$ and $E_{\rm p}$ are kinetic and potential energies) and three
masses for the molecular clouds of $M_{\rm g} = 10^4$, $4\times 10^5$,
and $10^6\,M_{\odot}$. The density of these molecular clouds are
$\rho_{\rm g} = 17$, $170$, and $1700$ cm$^{-3}$ (which corresponds to
1, 10, and 100 $M_{\odot}$pc$^{-3}$ assuming that the mean weight per
particle is $2.33m_{\rm H}$, respectively). The initial conditions are
summarized in Table \ref{tb:models_hydro}.

Once we chose the cloud mass and density, the radius ($R_{\rm g}$) and
the velocity dispersion in three dimensions ($\sigma_{\rm g}$) are
determined.  Some of our models (such as a models m1M-d1-s15,
m1M-d1-s16 and m1M-d1-s17, with $M_{\rm g}=10^6M_{\odot}$ and
$\rho_{\rm g}=17 {\rm cm}^{-3}$) roughly follow Larson's relation
\citep{1981MNRAS.194..809L},
\begin{eqnarray}
\sigma \sim \left( \frac{L}{1\mathrm{pc}}\right)^{0.5}(\mathrm{km}\,\mathrm{s}^{-1}),
\end{eqnarray}
where $\sigma$ is the velocity dispersion and $L$ is the size of the
cloud \citep{2004ApJ...615L..45H,2004RvMP...76..125M}.

\begin{figure*}
\begin{center}
\includegraphics[width=80mm]{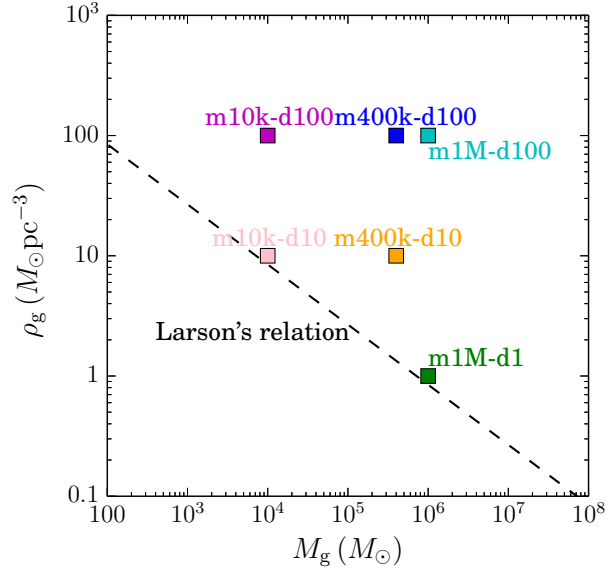}
\end{center}
\caption{Mass-density relation of our initial models (see
  Table\,\ref{fig:GMC_IC}). The dashed line indicates the mass-density
  relation from Larson's relation \citep{1981MNRAS.194..809L} for
  virialized cloud ($\sigma_{\rm g}^2=GM_{\rm g}/R_{\rm g}$).
\label{fig:GMC_IC}}
\end{figure*}

In Figure \ref{fig:GMC_IC} we present the distribution of mass and
density for the simulations listed in Table\,\ref{fig:GMC_IC}.  In
order to determine the mass of a molecular cloud that is consistent
with Larson's relation we adopt a velocity dispersion of $\sigma_{\rm
  g}\simeq \sqrt{GM_{\rm g}/R_{\rm g}}$.  Some models have initially a
higher velocity dispersion, which we motivate through cloud-cloud
collisions
\citep{2009ApJ...696L.115F,2013arXiv1306.2090F,2014ApJ...780...36F} or
to simulate molecular clouds in starburst galaxies.  We further
motivate and discuss on our choice of the initial conditions in
\S\,\ref{IC}.

\begin{table*}
\begin{center}
\caption{Initial conditions \label{tb:models_hydro} for the
  hydrodynamical simulations.  All models are super virial, with
  $|E_{\rm k}|/|E_{\rm p}| = 1$ }
\begin{tabular}{lcccccc}\hline \hline
Model  & Mass & Radius  & Density & Velocity dispersion & Initial free-fall time\\
      & $M_{\rm g} (M_{\odot})$ &$r_{\rm g}$ (pc)  & $\rho_{\rm g} ({\rm cm}^{-3})$ & $\sigma_{\rm g}$ (km\,s$^{-1}$) & $t_{\rm ff, i}$ (Myr)\\ \hline

m1M-d100-s7 & $1\times 10^6$ &  13.4  & $1.7\times10^3$  & 19.6 & 0.81 \\

m1M-d1-s15 & $1\times 10^6$ &  62  & 17  & 9.1  & 8.1 \\
m1M-d1-s16 & $1\times 10^6$ &  62  & 17  & 9.1  & 8.1  \\
m1M-d1-s17 & $1\times 10^6$ &  62  & 17  & 9.1  & 8.1  \\

m400k-d100-s1 & $4\times 10^5$ & 10 & $1.7\times10^3$  & 14.4 & 0.82\\
m400k-d100-s2 & $4\times 10^5$ & 10 & $1.7\times10^3$  & 14.4 & 0.82\\ 
m400k-d100-s3 & $4\times 10^5$ & 10 & $1.7\times10^3$  & 14.4 & 0.82\\

m400k-d10-s8 & $4\times 10^5$ & 21 & 170 & 9.9 & 2.5 \\
m400k-d10-s9 & $4\times 10^5$ & 21 & 170 & 9.9 & 2.5 \\

m10k-d100-s4 & $1\times 10^4$ & 2.87 & $1.7\times10^3$ & 4.2  & 0.81\\
m10k-d100-s5 & $1\times 10^4$ & 2.87 & $1.7\times10^3$ & 4.2  & 0.81\\
m10k-d100-s6 & $1\times 10^4$ & 2.87 & $1.7\times10^3$ & 4.2  & 0.81\\

m10k-d10-s11 & $1\times 10^4$ & 6.2 & 170 & 2.9 & 2.6  \\
m10k-d10-s12 & $1\times 10^4$ & 6.2 & 170 & 2.9 & 2.6  \\
m10k-d10-s13 & $1\times 10^4$ & 6.2 & 170 & 2.9 & 2.6  \\

\hline
\end{tabular}
\end{center}
\medskip
's' indicates the random seeds for the turbulence.
\end{table*}

\subsubsection{Smoothed Particle Hydrodynamics Simulations}

We perform hydrodynamical simulations using the SPH code {\tt Fi}
\citep{1989ApJS...70..419H,1997A&A...325..972G,2004A&A...422...55P,
  2005PhDT........17P} in the AMUSE framework.  
Our calculations have relatively low mass
resolution of $m=1M_{\odot}$ per particle. The gravitational softening
length during the hydrodynamical simulations is 0.1\,pc, and the SPH
softening length ($h$) is chosen such that $\rho_{\rm g} h^3 = m
N_{\rm nb}$ \citep{2002MNRAS.333..649S}. Here $N_{\rm nb} = 64$ is the
target number of neighbor particles. With the adopted isothermal gas
temperature of 30\,K we can resolve the Jeans instability down to $h
\sim 0.4$\,pc, which is smaller than the typical size of known embedded
clusters (1 pc) \citep{2003ARA&A..41...57L} but somewhat larger than
the observed typical width of gas filaments ($\sim 0.1$ pc)
\citep{2011A&A...529L...6A}. With these limitations, we obviously
cannot resolve the formation of individual stars, but we do resolve
dense gas clumps.  We think that the limited resolution of our
hydrodynamical simulations does not pose a serious problem, because we
are interested in the global dynamical structure of the molecular
cloud after only about an initial free-fall time scale, $t_{\rm ff,
  i}$ (see Table \ref{tb:models_hydro} for the free fall time scales
for each of the initial models). In fact, after $0.9 t_{\rm ff, i}$ we
stop the hydrodynamical simulation to analyze the resulting gas
distribution, initialize stars, and continue the simulations using a
gravitational $N$-body code.

\subsection{The star formation}

After stopping the hydrodynamical simulation (around $\sim 0.9t_{\rm
  ff, i}$) we replace some of the SPH particles with stellar
particles.  The selection of SPH particles is based, through the local
gas density $\rho$, on the local star formation efficiency (SFE)
$\epsilon_{\rm loc}$:
\begin{eqnarray}
  \epsilon_{\rm loc} = \alpha_{\rm sfe} 
                       \left( \frac{\rho}{100\,M_{\odot}{\rm pc}^{-3}}\right)^{0.5}.
\label{eq:eff}
\end{eqnarray}
Here $\alpha_{\rm sfe}$ is a free parameter in our simulations to
control the SFE.  the form of $\epsilon_{\rm loc}$ (Eq.\,\ref{eq:eff})
is motivated by the observations of individual molecular clouds for
which the star formation rate is argued to scales with local free-fall
time scale \citep{2012ApJ...745...69K,2013MNRAS.436.3167F}.

Here we adopt $\alpha_{\rm sfe}=0.02$, which reproduces the observed
global SFE across an entire molecular cloud of several per cent, but
also leads to a 10--30\,\% SFE in dense regions ($>1000 M_{\odot}{\rm
  cm}^{-3}$)
\citep{2003ARA&A..41...57L,2009ApJ...705..468H,2013ApJ...763...51F}.
In Table \ref{tb:models_Nbody} we present the global SFE ($\epsilon$)
and the SFE for the dense regions ($\epsilon_{\rm d}$) in our
simulations.

Depending on the local SFE we replace individual gas particles to
individual stellar particles conserving their positions and
velocities.  For each selected particle we assign a mass from the
Salpeter mass function \citep{1955ApJ...121..161S} between
$0.3\,M_{\odot}$\, and $100 M_{\odot}$, irrespective of the mass of its
parent SPH particle. The mean mass of the adopted mass function is
$1\,M_{\odot}$, which corresponds to the mass of individual SPH
particles. Mass in our simulations is therefore globally conserved,
but not locally.

\subsection{$N$-body simulations}

After the stellar particles are initialized (mass randomly from the
Salpeter mass function, and position and velocity from the parent SPH
particle), we remove the residual gas, leaving only the stellar
particles in the simulations. The instantaneous removal of the gas
has not a dramatic effect on the stellar distribution, because most
stars are formed in the densest regions where little low-density
(residual) gas is present. The gas that is insufficiently dense to
form stars tend to be enveloping the densest stellar conglomerates.

We now switch on the $N$-body code, for which we adopted the direct
sixth-order Hermite predictor-corrector scheme
\citep{2008NewA...13..498N} without gravitational softening and with
an accuracy parameter, $\eta=$0.1--0.25.  The total energy error over
the time span of the N-body simulations remained below $\sim 10^{-3}$.

The sizes of the stars we adopted from the zero-age main sequence
radii for solar metallicity stars \citet{2000MNRAS.315..543H}. We allow
stars to collide using the sticky sphere approach. New stellar
radii are assumed to be the zero-age main sequence radii for the 
new mass. Stellar mass-loss was incorporated only at the end of the main
sequence \citet{2000MNRAS.315..543H} \citep[see][for the
  details]{2009ApJ...695.1421F, 2013MNRAS.430.1018F}.

We did not perform the $N$-body simulations for models m10k-d10
($M_{\rm g}=10^4 M_{\odot}$ and $\rho_{\rm g}=10M_{\odot}{\rm pc}^{-3}
= 170 {\rm cm}^{-3}$), because the hydrodynamical simulations
resulted in less than 100 stars and we aim at $\gtrsim 100 M_{\odot}$
star clusters.  In these simulations even the densest regions were $<
1000 M_{\odot}{\rm pc}^{-3}$.

\begin{table*}
\begin{center}
\caption{Models\label{tb:models_Nbody} for $N$-body simulations}
\begin{tabular}{lcccccc}\hline \hline
Model  & Mass & $N$ of particles  & Virial ratio\tablenotemark{*} & SFE (Global) & SFE (Dense)\\
      & $M_{\rm s}(M_{\odot})$ &$N_{\rm s}$  & $|E_{\rm k}|/|E_{\rm p}|$ & $\epsilon$ & $\epsilon_{\rm d}$  \\ \hline

m1M-d100-s7 & $1.1\times 10^5$ & 109080 & 0.9 &  0.11 & 0.27\\

m1M-d1-s16 & $1.9\times10^4$ & 18760 & 0.50 &  0.019 & 0.63\\
m1M-d1-s16-t0.75 & $4.6\times10^3$ & 4566 & 19 & 0.0046 & 0.42\\
m1M-d1-s16-t0.65 & $3.9\times10^3$ & 3855 & 131 & 0.0039 & 0.083\\

m1M-d1-s15-t0.75 & $5.9\times10^3$ & 5902 & 1.6 &  0.0059 & 0.49\\
m1M-d1-s15-t0.65 & $4.0\times10^3$ & 3954 & 80  & 0.0040 & 0.12\\

m1M-d1-s17-t0.75 & $5.5\times10^3$ & 5506 & 6.4 & 0.0055 & 0.26\\
m1M-d1-s17-t0.65 & $4.3\times10^3$ & 4322 & 63  & 0.0043 & 0.088\\

m400k-d100-s1 & $3.2\times 10^4$ & 31895 & 1.3  & 0.078 & 0.22\\ 
m400k-d100-s2 & $2.3\times 10^4$ & 23273 & 4.2  & 0.057 & 0.16\\ 
m400k-d100-s3 & $4.3\times 10^4$ & 42596 & 0.43 & 0.096 & 0.25\\ 

m400k-d10-s8 &  $1.5\times 10^4$ & 14978 & 1.4 & 0.037 & 0.38\\
m400k-d10-s9 &  $2.8\times 10^4$ & 27891 & 0.41 & 0.068 & 0.39\\

m10k-d100-s4 & $4.1\times 10^2$ & 406 & 5.9 & 0.042 & 0.11\\
m10k-d100-s5 & $2.6\times 10^2$ & 256 & 7.4 & 0.027 & 0.079\\
m10k-d100-s6 & $2.5\times 10^2$ & 246 & 8.4 & 0.026 & 0.078\\

m10k-d10-s11 & 49 & 49 & - & 0.0049 & 0.00 \\
m10k-d10-s12 & 61 & 61 & - & 0.0061 & 0.00 \\
m10k-d10-s13 & 65 & 65 & - & 0.0065 & 0.00 \\

\hline
\end{tabular}
\end{center}
\medskip
's' indicates the random seeds for the turbulence.
\tablenotetext{*}{$|E_{\rm k}|$ and $|E_{\rm p}|$ are the total
  kinetic and potential energies of the entire stellar system,
  respectively. For virialized systems the virial ratio equals
  0.5. For models m10k-d10 we did not perform $N$-body simulations,
  and therefore their virial ratio is not calculated.}
\end{table*}

\section{Results}\label{Sect:Results}

\subsection{Formation of Embedded, Classical Open, and Young Massive Clusters}

\begin{figure*}
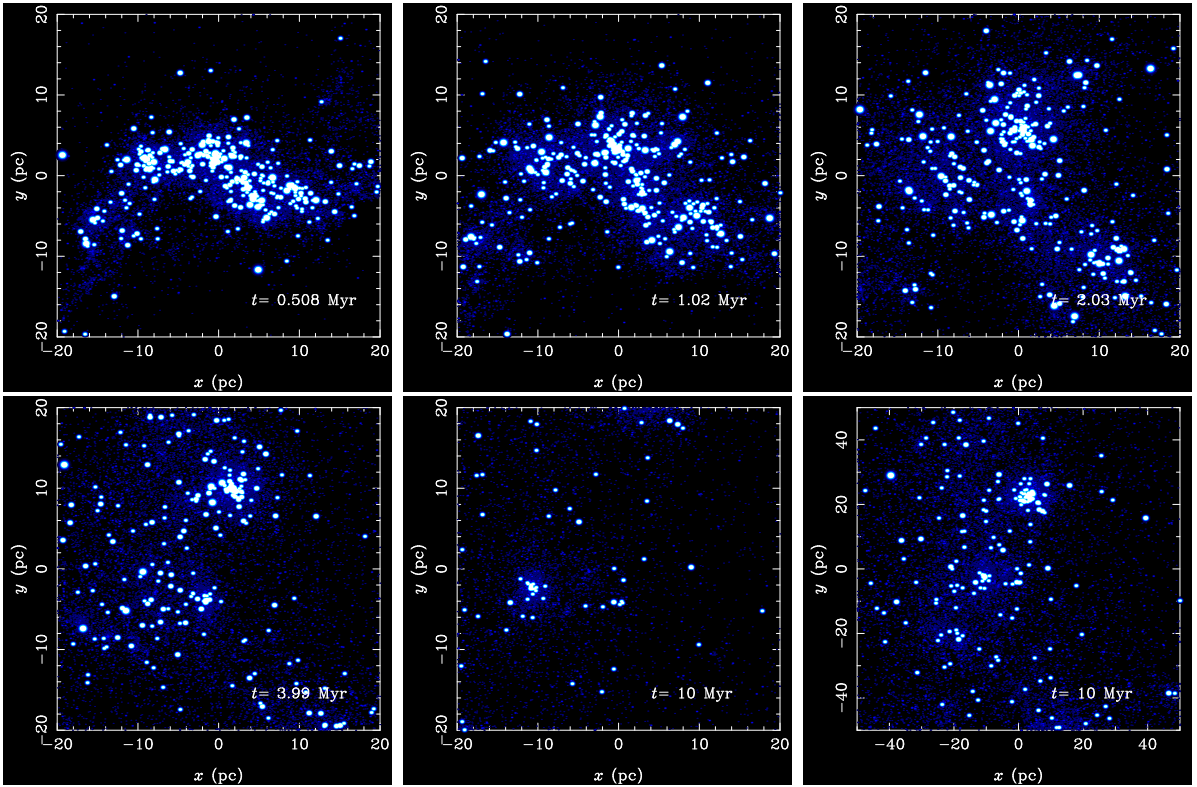

\begin{center}
\includegraphics[width=52mm]{f2a.eps}
\includegraphics[width=52mm]{f2b.eps}
\includegraphics[width=52mm]{f2c.eps}
\includegraphics[width=52mm]{f2d.eps}
\includegraphics[width=52mm]{f2e.eps}
\includegraphics[width=52mm]{f2f.eps}
\end{center}
\caption{Snapshots of model m400k-d100-s3.  The size of the dots
  indicate the masses of the stars: $8<m/M_\odot<16$ for the small
  dots, $16<m/M_\odot<40$ for middle sized and $m>40 M_{\odot}$ for
  the largest dots.  Stars with a mass $m<8 M_{\odot}$ are plotted as
  small blue dots.
\label{fig:snapshots}}
\end{figure*}

The $N$-body simulations are started at what we will call $t=0$\,Myr.
The initial distribution of stars follows the distribution of the
densest regions in the turbulent molecular cloud.  In Figure
\ref{fig:snapshots} we present a time series of snapshots of model
m400k-d100-s3.  The entire system continuously expands because not all
stars are bound after gas expulsion. The distribution of stars is
clumpy and it takes a few Myr before the stars assemble in a more
coherent aggregate.

We interrupt the simulations twice, at $t=2$ and at $t=10$\,Myr,
in order to analyze the stellar distribution, and detect clustered
aggregates. Clumps are found in these snapshots by means of {\tt HOP}
\citep{1998ApJ...498..137E} in AMUSE, using an outer cut-off density of
$\rho_{\rm out} = 4.5M_{\rm s}/(4\pi R_{\rm h}^{3})$ (three times the
half-mass density of the entire stellar system, $\rho_{\rm h}=M_{\rm
  s}/(8\pi R_{\rm h}^3)$, where $M_{\rm s}$ is the total stellar mass
and $R_{\rm h}$ is the half-mass radius of the entire distribution of
the stars), a saddle-point density threshold ($\rho_{\rm saddle} =
8\rho_{\rm out}$) and the peak density threshold ($\rho_{\rm peak} =
10\rho_{\rm out}$) and the number of particles for neighbor search
($N_{\rm dense}$) as well as the number of particles to calculate the
local density ($N_{\rm hop}$) are set to be 64.  The number of
neighbors is used to determine which two groups merge $N_{\rm
  merge}=4$.  With these settings the detection limit of the clump
mass is $\sim 100M_{\odot}$.  Sometimes {\tt HOP} identifies multiple
clumps as one, but by applying the method repeatedly we can separate
those again.  For this iterative procedure we adopt $\rho_{\rm out} =
\rho_{\rm h, c}$, where $\rho_{\rm h, c}$ is the half-mass density
of a detected clump.  
We continue this procedure until $\rho_{\rm h, c} \gtrsim 100\rho_{\rm h}$, 
after which the clumps are so dense compared to the background 
that they do not separate anymore in substructures
\citep[see][for the details]{2015PASJ..tmp..163F}.

\begin{figure*}
\begin{center}
\includegraphics[width=80mm]{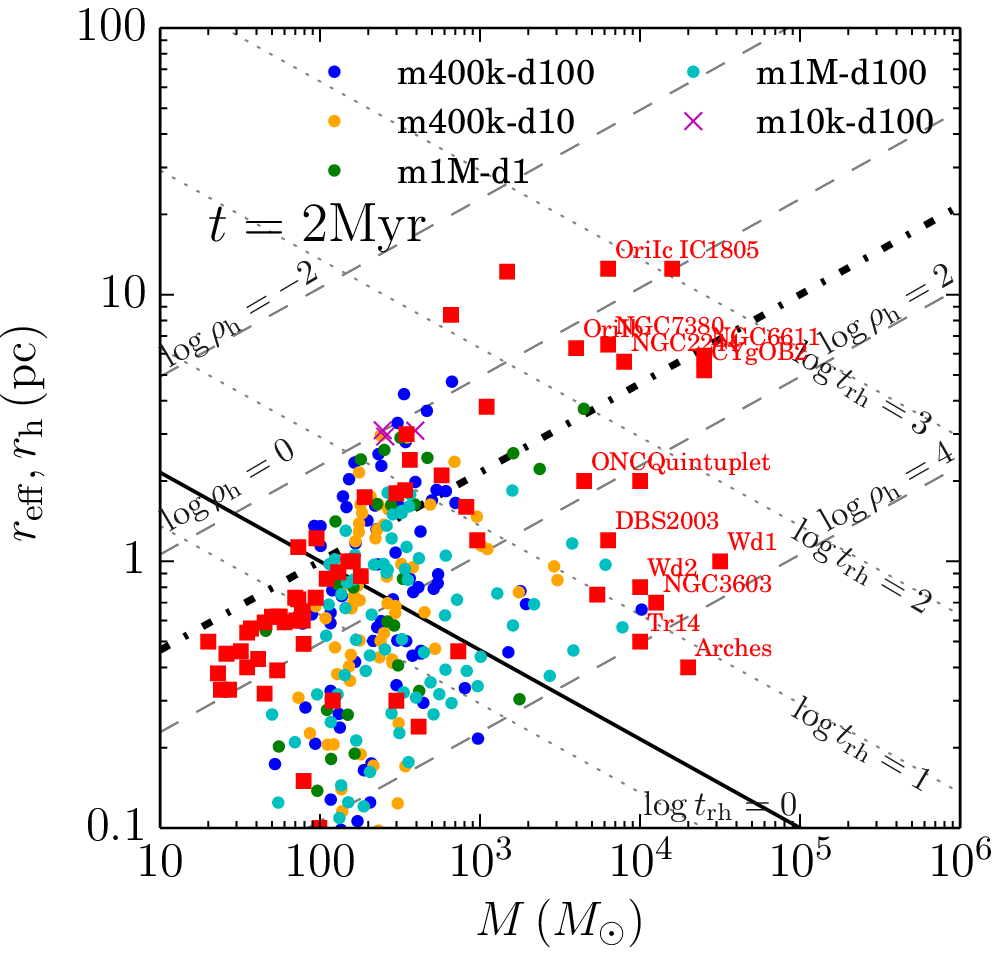}
\includegraphics[width=80mm]{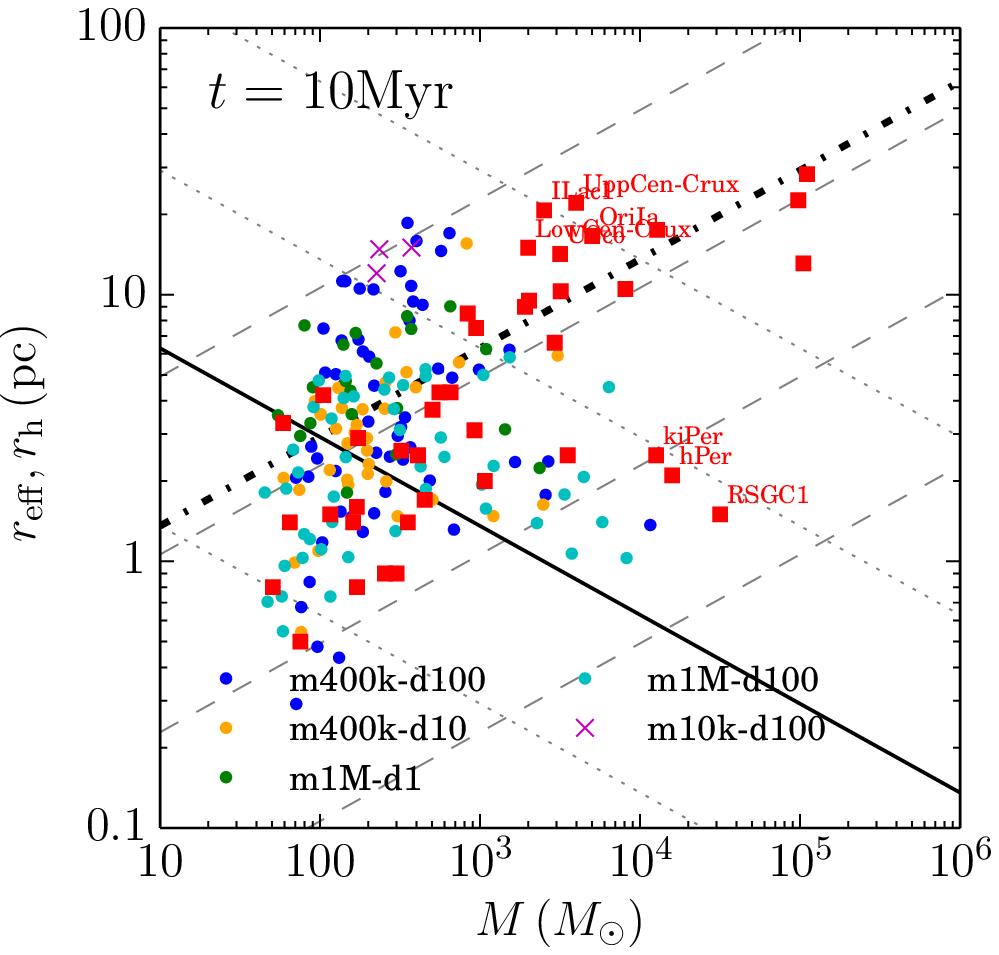}
\end{center}
\caption{Mass-radius diagram of observed and simulated clusters at
  $t=2$ and 10 Myr since the start of the $N$-body simulations.
  Colored dots are clusters obtained from simulations using a clump
  finding method.  Crosses indicate the median radius and the total
  mass of the entire stellar system rather than the detected
  individual clusters.  Red squares indicate observed clusters with an
  age of 1--5\,Myr (left) and 5--15\,Myr (right). Data are from
  \citet{2008A&A...487..557P,
    2009AJ....137.4777W,2003ApJ...593.1093L,2006AJ....132.2296A,
    2009A&A...504..461F,2006ApJ...646.1215L,2008AJ....135..966F,
    2011MNRAS.414.3769B,1997AJ....113.1788H,1997MNRAS.286..538D,
    1991AJ....102.1108H,2010ARA&A..48..431P}.  Observed clusters with
  names are the clusters listed in \citet{2009A&A...498L..37P} and
  \citet{2010ARA&A..48..431P}.  Black thick solid and dash-dotted
  lines indicate the line at which the relaxation time and the
  dynamical time are equal to the age of the stellar populations.
  Gray dashed lines indicate the half-mass density of 0.01, 1, 100,
  and $10^4 M_{\odot} {\rm pc}^{-3}$, and gray dotted lines indicate
  the half-mass relaxation time of 1000, 100, 10, and 1 Myr from top
  to bottom.  We used the median radius for the observed leaky
  clusters \citep{2007AJ....133.1092W,2009A&A...498L..37P}.
\label{fig:MR1}}
\end{figure*}

In Figure \ref{fig:MR1} we present the mass and half-mass radius of
the star clusters obtained from our simulations at $t=2$ and at
$t=10$\,Myr.  For comparison, we added a number of observed open
clusters (classical open, embedded, young massive, and leaky clusters)
to the same diagram.  The majority of the identified clusters have
masses and radii consistent with those of classical open clusters
\citep{2008A&A...487..557P} \citep[see also][]{2015PASJ..tmp..163F}
and of known embedded clusters \citep{2003ARA&A..41...57L}.  The densest
initial molecular clouds (m1M-d100, m400k-d100, and m400k-d10) tend to
to form massive compact clusters, similar to young massive clusters.
Such compact clusters do not form in the less dense or less massive 
molecular clouds (such as m1M-d1 or m10k-d100).

When observing the 10-Myr-old stellar conglomerates from a distance,
they tend to blend in a single star forming region with an average
density of $\sim 0.01M_{\odot}{\rm pc}^{-3}$, which is comparable to
the mean field density in solar neighborhood
\citep{2000MNRAS.313..209H}.  Such conglomerates may remain
unrecognizable as a cluster system.
For those simulations in which no clumps are detected down to a limit
of $100\,M_\odot$, we adopt the median distance of the stars from the
cluster center.

The masses and half-mass radii of the clusters in our simulations
mainly resemble the populations of observed embedded and classical 
open clusters.
This result appears to be independent of the initial molecular-cloud
density.   Embedded and classical open clusters cluster
around the point where the cluster age ($t_{\rm age}$) equals the
dynamical time ($t_{\rm dyn}$) and the half-mass relaxation time
($t_{\rm rh}$) \citet[see also][]{2015PASJ..tmp..163F}.

Here the dynamical time and the half-mass relaxation time are written
as
\begin{eqnarray}
t_{\rm dyn} \sim 2\times 10^4\left( \frac{M}{10^6M_{\odot}}
               \right)^{-1/2} \left(\frac{r_{\rm h}}{1{\rm pc}}\right)^{3/2}
                     {\rm year}
\label{eq:t_dyn}
\end{eqnarray}
and
\begin{eqnarray}
t_{\rm rh} \sim 2\times 10^8 \left( \frac{M}{10^6M_{\odot}} \right)^{1/2}
              \left( \frac{r_{\rm h}}{1{\rm pc}}\right)^{3/2} {\rm year},
\label{eq:t_rh}
\end{eqnarray}
respectively \citep{2010ARA&A..48..431P}, where $M$ is the cluster
mass, and $r_{\rm h}$ is the half-mass radius.  For clarity we assumed
that the virial radius of star clusters is comparable to the half-mass
radius and that the mean stellar mass is $1\,M_{\odot}$ (as is the
case in our simulations).  In Figure \ref{fig:MR1} we present lines
on which the relaxation (black full) and dynamical (black dash-dotted) 
times are equal to the age of the clusters, respectively.  
Both lines move as well as all the symbols upward with time.

For the formation of young massive clusters, we find that a dense massive
molecular cloud is necessary.  The densities required to form such
massive clusters exceed the density expected by Larson's relation; the
velocity dispersion necessary for the formation of young massive
clusters is too high.  Such an initial high density may be realized by
cloud-cloud collisions \citep{2014ApJ...780...36F}.
The velocity dispersion of our dense model $\sim 20$ km\,s$^{-1}$,
which is comparable to the typical relative velocity of molecular
clouds associated with young massive clusters such as NGC 3603 and
Westerlund 2. For these clusters a collision between two molecular
clouds was considered to trigger their formation
\citep{2009ApJ...696L.115F,2010ApJ...709..975O,2013arXiv1306.2090F,
  2014ApJ...780...36F}, which is consistent with our findings here.

For forming a star cluster in our simulations, the molecular cloud
must be compressive (a high velocity dispersion due to a high
density), which is consistent with observations
\citep{2007ARA&A..45..481Z}.  From various initial conditions, we find
that star clusters similar to open, known embedded, and young massive
clusters form in these simulations, but leaky clusters
($M\sim10^4M_{\odot}$ and $r_{\rm h}\sim 10$ pc) must form from
different initial conditions.  We discuss the formation of leaky
clusters in the following section
(\S\,\ref{Sect:LeakyClusterFormation}).

\subsection{Formation of Leaky Clusters}\label{Sect:LeakyClusterFormation}

In the previous section, we show that known embedded, classical open, and
young massive clusters form from turbulent molecular clouds, but no
leaky cluster is found in our simulations. In this section we address
the question: how do leaky clusters form?  Is the formation process
different from the other clusters?

\citet{2011A&A...536A..90P} proposed that observed embedded clusters grow in
mass and size due to star formation and become leaky clusters as a
result of the expulsion of the residual gas.  This scenario was later
explored and the evolutionary tracks of such a cluster on the
mass-radius diagram were suggested
\citep{2013A&A...549A.132P,2013A&A...559A..38P,2014ApJ...794..147P}.
\citet{2010ARA&A..48..431P}, however, classified the leaky clusters as
OB associations. We here do not discuss if the leaky clusters are
associations or clusters, but treat both leaky clusters and
associations as less dense clustered systems.

\begin{figure*}
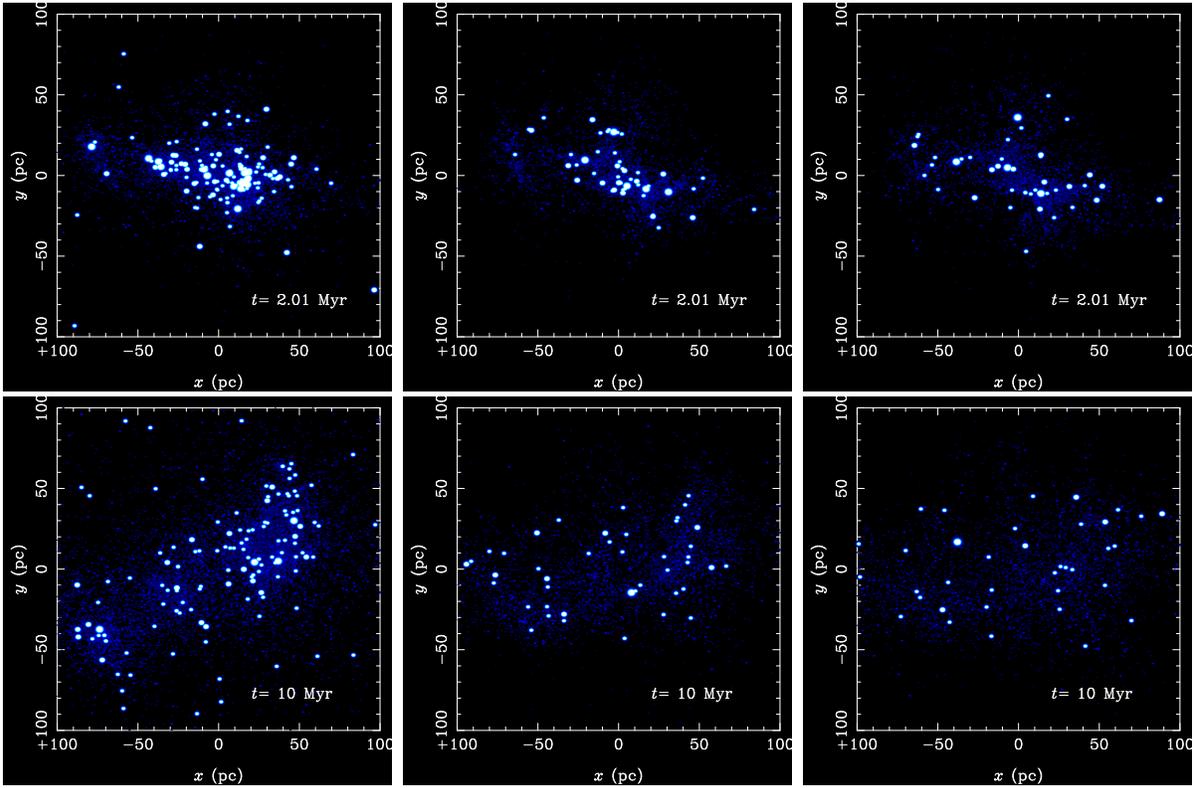

\begin{center}
\includegraphics[width=52mm]{f4a.eps}
\includegraphics[width=52mm]{f4b.eps}
\includegraphics[width=52mm]{f4c.eps}
\\
\includegraphics[width=52mm]{f4d.eps}
\includegraphics[width=52mm]{f4e.eps}
\includegraphics[width=52mm]{f4f.eps}
\end{center}
\caption{Snapshots at $t=2$ (top) and 10 (bottom) Myr for model d1-1M,
  but for different timing of gas removal. $t=0.9, 0.75$, and
  $0.65t_{\rm ff, i}$ (models m1M-d1-s16, m1M-d1-s16-t0.75, and
  m1M-d1-s16-t0.65) from left to right.
\label{fig:snap_d1-1M}}
\end{figure*}

We consider leaky clusters (and also OB associations) to form clumpy
but that they lose this structure in the early dynamical evolution,
contrary to the arguments in \citet{2011A&A...536A..90P}.  We support
our argument with the simulation model m1M-d1-s16 (see the left panels
in Figure \ref{fig:snap_d1-1M}).  This simulation started with a
spherical molecular cloud that collapsed asymmetrically due to the
turbulence velocity field.  Stars formed mainly in the densest regions
which result in the stellar distribution being elongated and clumpy.

\begin{figure*}
\begin{center}
\includegraphics[width=80mm]{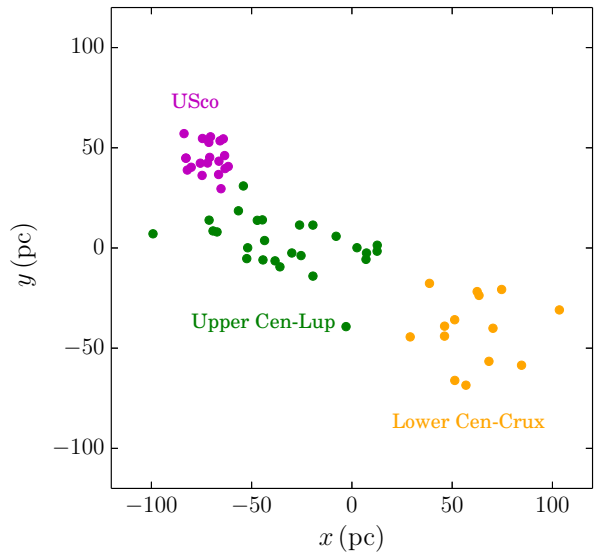}
\end{center}
\caption{Positions of B-type stars which belong to USco (magenta),
Upper Cen-Lup (green), and Lower Cen-Crux (orange). 
Data is from \citet{2007AJ....133.1092W}. We assume 140 pc as the 
distance \citep{2007AJ....133.1092W}.
\label{fig:UCL_LCC}}
\end{figure*}

After the residual gas has been removed, the clusters tend to be super
virial, and some stars escape right away (see the virial ratio given in
Table \ref{tb:models_Nbody}). As a consequence, the entirely stellar
distribution expands with time.  At an age of $t=10$\,Myr the density
of the environment has decreased substantially, and the spatial
distribution of the stars resembles leaky clusters and OB
associations.  In Figure \ref{fig:UCL_LCC} we present the spatial
distribution of O and B spectral-type stars in the association
Scorpius OB2 (Sco OB2), which can be compared with our simulations in
Figure \ref{fig:snap_d1-1M}.

Sco OB2 is composed of three subgroups; Upper Scorpius (USco), Upper
Centaurus-Lupus (Upper Cen-Lup), and Lower Centaurus-Crux (Lower
Cen-Crux) \citep{2007AJ....133.1092W}.  These subgroups are listed in
\citet{2009A&A...498L..37P} as leaky clusters, and as associations in
\citet{2010ARA&A..48..431P}.  They are all located at similar
distances from the sun, at 145\,pc, 142, and 118\,pc, respectively
\citep{2007AJ....133.1092W}, and therefore they are considered to be a
system.  The distribution of massive (O and B) stars in Sco OB2 is
very similar to the distribution of massive
stars in model m1M-d1-s16 at an age of 10 Myr.

\begin{figure*}
\begin{center}
\includegraphics[width=80mm]{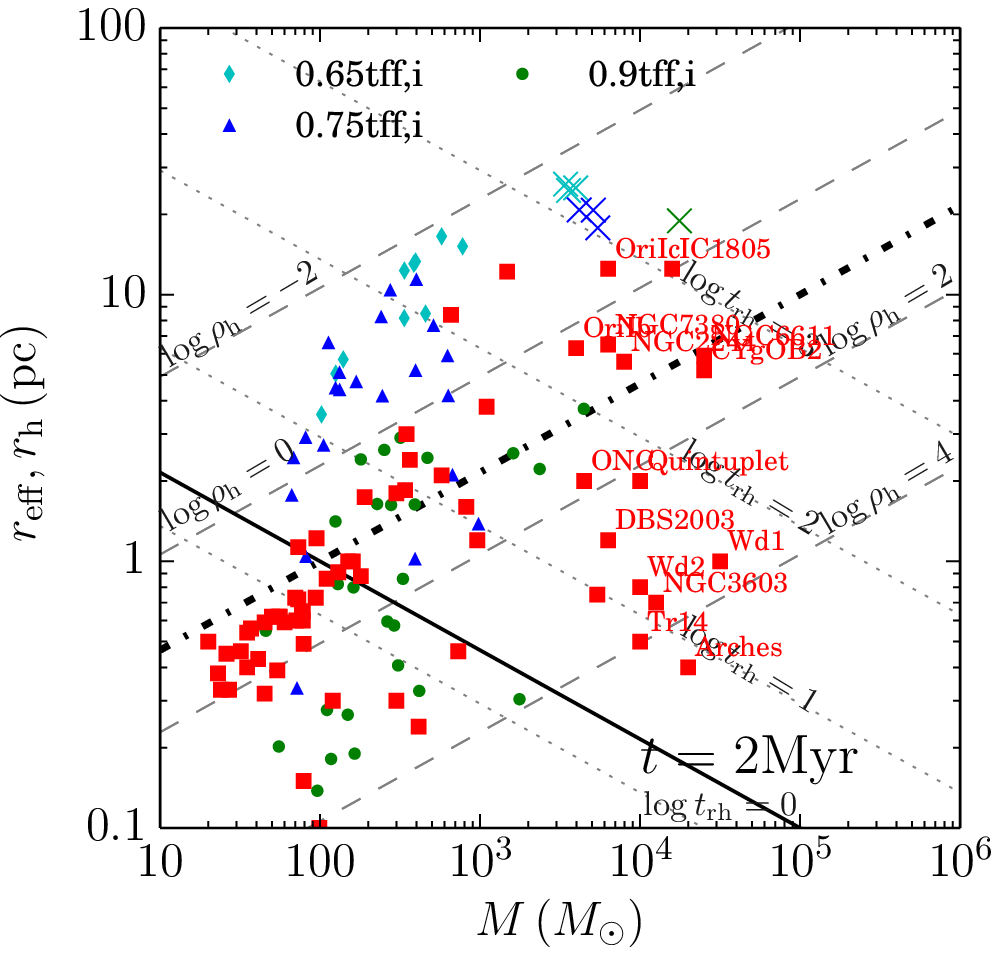}
\includegraphics[width=80mm]{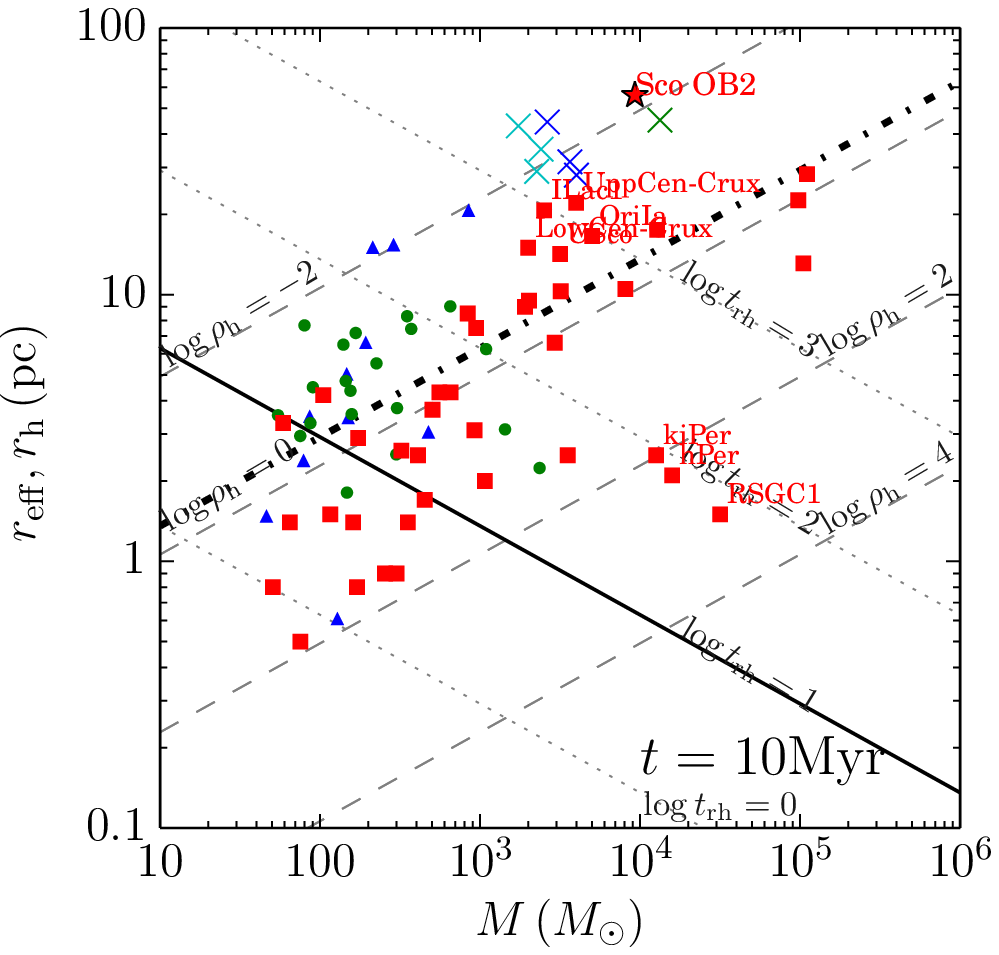}
\end{center}
\caption{Mass-radius diagram of detected clusters for models
  m1M-d1-s16 (green dots), m1M-d1-s16-t0.75 (blue triangles), and
  m1M-d1-s16-t0.65 (cyan diamonds) at $t=2$ and 10 Myr.  Red squares
  indicate observed young clusters; clusters with an age of 1--5 Myr
  and 5--15 Myr are plotted in top and bottom panels,
  respectively. The data for the observed clusters are the same as
  those in Figure \ref{fig:MR1}.  The mass
  and half-mass radii of simulations, interpreted as unresolved
  clusters, are shown as crosses; each cross represents a single
  simulation.
\label{fig:MRD}}
\end{figure*}

In figure \ref{fig:MRD} we present the result of our clump finding
analysis for model m1M-d1-s16 at 2\,Myr and at 10 Myr. At $t=2$\,Myr
we detected $\sim 20$ clusters that are similar to observed embedded
star clusters. At $t=10$ Myr no clear massive clusters remain visible
in the snapshot (see Figure \ref{fig:snap_d1-1M}), although we still
detected several classic open cluster-like structures; in the epoch
between 2\,Myr to 10\,Myr, the stellar distribution has dispersed.

When interpreting the entire system in each simulation as a single
association, the mass and radius are very similar to those of observed
leaky clusters and OB associations. In figure \ref{fig:MRD} we present
these as crosses (to the top of the panels at 2\,Myr and 10\,Myr).
Model m1M-d1-s16 has a similar appearance and dynamical structure as
the Sco OB2 system, rather than the individual sub-clusters USco,
Upper Cen-Lups and Lower Cen-Crux.  In this analysis we excluded
single stars (those with a local density $\rho_6 <10^{-3}M_{\odot}{\rm
  pc}^{-3}$) which is more than an order of magnitude lower than the
mean density of the solar neighborhood ($\rho_6$ here is the density
measure within the 6 nearest neighbors).

We still detected clumps consistent with open clusters in model
m1M-d1-s16.  These clumps are the result of the clumpiness of
molecular clouds at a time when we stop the hydrodynamical simulations
(at $\sim 0.9t_{\rm ff, i}$).  In observed star forming regions,
however, stars appear to form when the local density exceeds some
threshold density for self-gravitating clouds of $\sim 10^3{\rm
  cm}^{-3}$ \citep{2007ARA&A..45..565M}, and feedback starts to
dominate the hydrodynamics as soon as the first massive star forms,
which may happened well before a free-fall time scale.
The free-fall time scale of model m1M-d1-s16 is $\sim 8$\,Myr, which
is considerably longer than the formation time for massive stars
($\sim1$\,Myr) \citep{2007ARA&A..45..565M}. In such a region, where the
star forming time scale is considerably smaller than the free-fall
time scale of the entire molecular cloud, stellar feedback is expected 
to terminate the star formation before the molecular cloud fully collapses.
This would result in a less clumpy stellar distribution. 

Unfortunately in our simulations, we cannot take such gradual star
formation and feedback processes into account, although they have been
addressed with the AMUSE framework by \cite{2012MNRAS.420.1503P}. In
order to mimic the early star formation process, we experimented with
stopping the hydrodynamical simulations at an earlier epoch and
replace the gas particles with stellar particles.

As in our previous simulations, we assumed that the feedback
terminates star formation which causes the residual gas to be ejected
instantaneously.  We stop the hydrodynamical simulation for model
m1M-d1-s16 at $t=0.65 t_{\rm ff, i}$ and $0.75 t_{\rm ff, i}$ (5.3 and 6.2 Myr,
respectively), and replace gas particles to stellar particles using
the same way as for model m1M-d1-s16, i.e., assuming a local star
formation efficiency given by equation (\ref{eq:eff}) and the same
values for $\alpha _{\rm sfe}=0.02$. The numbers of stars that form
using this procedure decrease considerably, and the resulting virial
ratio of the stellar system increases.  We also run the same initial
conditions but with different random seeds (m1M-d1-s15 and m1M-d1-s17).
In Table \ref{tb:models_Nbody}
we present some global parameters for these models.

Snapshots of these models (m1M-d1-s16-t0.75 and m1M-d1-s16-t0.65) are
shown in the middle and right panels of Figure
\ref{fig:snap_d1-1M}. The distribution of massive stars is less clumpy
compared with that of model m1M-d1-s16 (standard model, in which the
hydrodynamical simulation is stopped at $0.9t_{\rm ff}$; see the left
panels of Figure \ref{fig:snap_d1-1M}).  We also apply the clump
finding algorithm to these models, and the results of which are shown in
Figure \ref{fig:MRD}. At $t=2$Myr, several clumps are detected in both
models, but they are less dense compared with those detected in our
standard model. In model m1M-d1-s16-t0.65 in particular, the density of
the detected clumps is only slightly elevated compared to the
background density in the solar neighborhood ($0.01M_{\odot}{\rm
  pc}^{-3}$) \citep{2000MNRAS.313..209H}, and these clumps may
therefore not be recognized as clusters.  In model m1M-d1-s16-t0.75 some
clumps which resemble open clusters are still detected at $t=10$ Myr,
but none in model m1M-d1-s16-t0.65.  If we treat the entire system as a
one cluster, the masses are similar to those of leaky clusters and
associations, even though the size remains larger by about a factor of
two.  On Figure \ref{fig:MRD} we present the resulting clusters with a
mass and half-mass radius in stars with $\rho_{6}>10^{-3}M_{\odot}{\rm
  pc^{-3}}$.

Our assumption that star formation terminates instantaneously
throughout the system after about one free-fall time of the
molecular cloud probably overestimates the effect of the feedback
considerably. In observed star-forming regions the feedback from
massive stars tend to limit star formation locally, but may not affect
the entire ($\sim100$ pc across) star forming region.  In the
simulations of \citet{2012MNRAS.420.1503P}, the wind of one massive
$\sim 30M_{\odot}$ star blows the residual gas from the clustered
environment in a couple of Myr, which is much longer than adopted in
our simulations.

If star formation proceeds as clumpy as simulated here, the feedback
is even more localized, which will result in a considerable age spread
among subgroups.  Our simulations would then be representative for the
formation of cluster complexes such as USco, Upper Cen-Lups, and Lower
Cen-Crux, or OB association such as Sco OB2.  The ages of these three
subgroups are slightly different each other; 14--15, 11--12, and 5--6
Myr for Upper Cen-Lup, Lower Cen-Crux, and USco, respectively
\citep{2007AJ....133.1092W}.  
If we could assume local feedback processes, an
association (or leaky clusters) similar to Sco OB2 might form from an
initial condition such as models m1M-d1.  
Less dense clusters tend to
have a wider age spreads \citep{2014ApJ...791..132P}, which is also
consistent with our simulations.  We therefore argue that the
ancestors of associations are conglomerates of denser embedded
clusters.  We detect these as an environment with multiple low-mass
but rather dense clusters that disperse in time. The evaporation of
these clusters is driven by relaxation and feedback, and this makes
them resemble associations.

\section{Initial Conditions of Molecular Clouds}
\label{IC}

In the previous section, we showed that our dense models tend to form
young massive clusters and that less dense models lead to leaky
clusters as well as known embedded and classic open clusters.  The types of
the resulting star cluster is sensitive to the initial conditions of
the parental molecular clouds.  In this section, we compare our
initial conditions with observed molecular clouds and discuss a model
for the formation of clusters in the Milky Way and other nearby
galaxies.

\begin{figure*}
\begin{center}
\includegraphics[width=80mm]{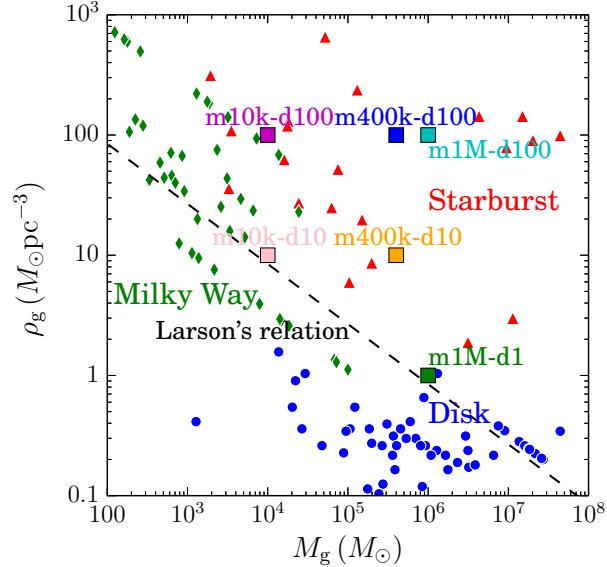}
\end{center}
\caption{Mass-density relation of observed molecular clouds. Green diamonds indicate individual molecular clouds in the Milky Way galaxy. Blue circles and red triangles are for molecular clouds typical in individual local disk and starburst galaxies, respectively. Each point indicates one galaxy. The data is from \citet{2012ApJ...745...69K}. Color squares indicate our initial conditions which are the same as those shown in Figure \ref{fig:GMC_IC}.  Dashed line indicates
the mass-density relation following Larson's law for virialized
cloud ($\sigma_{\rm g}^2=GM_{\rm g}/R_{\rm g}$).
\label{fig:IC}}
\end{figure*}

In Figure \ref{fig:IC}, we present the mass and density of individual
molecular clouds observed in the Milky Way and those estimated for
local disk and starburst galaxies \citep{2012ApJ...745...69K}. We also
show the initial conditions of our simulations.
The dashed line in the Figure \ref{fig:IC} indicates the Larson's
relation.  In order to estimate the mass of molecular clouds following
Larson's law, we assume that the molecular clouds are in virial
equilibrium (i.e., they satisfy $\sigma_{\rm g}^2=GM_{\rm g}/r_{\rm
  g}$, where $\sigma_{\rm g}$, $M_{\rm g}$, and $r_{\rm g}$ are the
velocity dispersion, mass, and radius of the molecular clouds, respectively).
Observed molecular clouds, however, are not necessarily virialized.

Molecular clouds in the Milky Way tend to follow Larson's relation, but
with a large scatter of the density.  On the other hand,
not all of our initial conditions are consistent with the mass and
density of molecular clouds observed in the Milky Way.  Models
m10k-d100 ($10^4M_{\odot}$ and $100M_{\odot}{\rm
  pc}^{-3}\simeq1700$cm$^{-3}$) and m10k-d10 ($10^4M_{\odot}$ and
$10M_{\odot}{\rm pc}^{-3}\simeq170$cm$^{-3}$), for example, are
initially indistinguishable from typical molecular clouds in the
Milky Way.

As we described in section~\ref{Sect:Results}, the number of stars
formed in model m10k-d10 was too small (fewer than 100 stars) to be
recognized as a cluster in our analysis. Model m10k-d100 produces a
sufficiently large number of stars but does not form a recognizable
cluster after 2\,Myr.  If we treat the entire region of this model as
a cluster conglomerate, the mass and radius is similar to that of an
open cluster.  From this, we conclude that the molecular clouds
typical in the Milky Way tend to form classical open clusters, but
that they are insufficiently massive and dense to form massive star
clusters.

Model m1M-d1 ($10^6M_{\odot}$ and $1M_{\odot}{\rm pc}^{-3}$)
represents the most massive molecular cloud in the Milky Way
\citep{2011ApJ...729..133M}, and it follows Larson's relation. This initial
condition results in several embedded cluster cores, that eventually
evolve to a conglomerate of associations.

The initial conditions which tend to form young massive clusters are
considerably denser than the molecular clouds observed in the Milky
Way (see Figure \ref{fig:IC}).  To form a young massive clusters in
our simulations a mass of at least several $10^5M_{\odot}$ and a mean
density of $10M_{\odot}{\rm pc}^{-3}$ (170\,cm$^{-3}$) is required.
Such initial conditions are common in local starburst galaxies, but
very rare in the Milky Way.

In Figure \ref{fig:IC} we present the estimated mass and molecular
clouds density typical for local starburst and disk galaxies.  This
data is obtained from \citet{2012ApJ...745...69K}.  We calculated the
masses and densities for these molecular clouds from the free-fall
time scale provided by \citet{2012ApJ...745...69K} using the observed
surface gas densities ($\Sigma_{\rm g}$). In
\citet{2012ApJ...745...69K} they considered two rather distinct
regimes of molecular clouds; these are the molecular cloud regime and
the Toomre regime.  The molecular cloud regime is expected to be
common in local disk galaxies.  The molecular clouds are decoupled
from their surrounding interstellar medium, and as a result
self-gravitating \citep{2012ApJ...745...69K}.  The Toomre regime is
common in starburst galaxies.  In this case the interstellar medium is
highly turbulent and therefore the free-fall time scale of the
molecular clouds should be estimated using the mid-plane pressure in
the galactic disks \citep[see][for the details]{2012ApJ...745...69K}.

Following the description of \citet{2012ApJ...745...69K}, we estimate
the typical mass of molecular clouds for each galaxy listed in
\citet{2012ApJ...745...69K}.  We take the smaller free-fall time scale for
the molecular cloud and Toomre regimes ($t_{\rm ff, GMC}$ and $t_{\rm
  ff, T}$, respectively) as the free-fall time scale ($t_{\rm ff}$),
which is consistent with \citet{2012ApJ...745...69K}.  We calculate the density through
the free-fall time scale using:
\begin{eqnarray}
\rho_{\rm g} = \frac{3\pi}{32 G t_{\rm ff}^2}. \label{eq:tff}
\end{eqnarray}
In the Toomre regime, the mid-plane pressure in the disk of surface
gas density $\Sigma_{\rm g}$ is
\begin{eqnarray}
P = \rho_{\rm g,T} \sigma_{\rm g}^2 = \phi_{P}\frac{\pi}{2}G\Sigma_{\rm g}^2.
\label{eq:Toomre}
\end{eqnarray}
Here $\rho_{\rm g,T}$ is the molecular cloud density in the Toomre
regime, $\sigma_{\rm g}$ is the velocity dispersion of the gas and
$\phi_{P}$ is a dimensionless factor \citep{2012ApJ...745...69K}. The
Toomre $Q$ for the gas is written as
\begin{eqnarray}
Q = \frac{\sqrt{2(\beta+1)}\sigma_{\rm g}\Omega}{\pi G\Sigma_{\rm g}}.
\label{eq:Q}
\end{eqnarray}
Here $\beta$ is the logarithmic index of the rotation curve ($\beta=0$
for a flat rotation curve, whereas for solid-body rotation $\beta=1$),
$\Omega = 2\pi /t_{\rm orb}$ ($t_{\rm orb}$ is the galactic orbital
period) is the angular velocity of galactic rotation \citep[see
  also][]{2005ApJ...630..250K}.  From these equations, the density of
the molecular cloud becomes
\begin{eqnarray}
\rho_{\rm g, T} \simeq \frac{(\beta+1)\phi_{P}\Omega^2}{\pi G Q^2}.
\label{eq:rho_GMC}
\end{eqnarray}
Here we adopt $Q\sim1$ and $\beta=0$ following
\citet{2012ApJ...745...69K}.  If we assume that the cloud is
virialized ---i.e., $\sigma_{\rm g}^2\sim GM_{\rm g,T}/r_{\rm g}$,
where $M_{\rm g,T}$ and $r_{\rm g}$ are the mass and radius of the
cloud--- from equation (\ref{eq:Q}) and $\rho_{\rm g,T} = 3M_{\rm
  g,T}/(4\pi r_{\rm g}^3)$ we can estimate the cloud mass using:
\begin{eqnarray}
M_{\rm g,T} \sim \frac{1}{32} \sqrt{\frac{3}{\pi}} G^{3/2} \Sigma_{\rm
  g} ^3 t_{\rm orb}^3 \rho_{\rm g}^{-1/2}.
\label{eq:M_GMC}
\end{eqnarray}
Because for each galaxy $t_{\rm orb}$ and $\Sigma_{\rm g}$ are given
in \citet{2012ApJ...745...69K}, we can estimate the cloud density and
mass from equation (\ref{eq:rho_GMC}) and (\ref{eq:M_GMC}).  Here we
adopt $\phi_{P}\simeq3$, following \citet{2012ApJ...745...69K}.

For the molecular cloud regime (i.e., $t_{\rm ff, GMC}<t_{\rm ff,
  T}$), the mass is estimated as follows.  The mass of molecular
clouds is estimated by the two-dimensional Jeans mass in galactic
disks
\citep{2002ApJ...570..132K,2007ARA&A..45..565M,2011ApJ...727...88C},
which is given by
\begin{eqnarray}
M_{\rm g,GMC}=\frac{\sigma_{\rm g}^4}{G^2\Sigma _{\rm g}},
\label{eq:M_GMC2}
\end{eqnarray}
\citep[see equation\,(3) of][]{2012ApJ...745...69K}. Since the mass and
density of molecular clouds in the molecular cloud regime are written
as $M_{\rm g,GMC}=\pi r_{\rm g}^{2}\Sigma_{\rm GMC}$ and $\rho_{\rm
  g,GMC}=(3/4\pi)M_{\rm g,GMC}r_{\rm g}^{-3}$, where $\Sigma_{\rm
  GMC}$ is the surface density of molecular clouds, and using
equation (\ref{eq:M_GMC2}) we can calculate the density of molecular
clouds with \citep[equation (4) in][]{2012ApJ...745...69K}:
\begin{eqnarray}
\rho_{\rm g,GMC} = \frac{3\sqrt{\pi}}{4}\frac{G\sqrt{\Sigma _{\rm GMC}^3\Sigma_{\rm g}}}{\sigma_{\rm g}^2}.
\label{eq:rho_GMC2}
\end{eqnarray}
Here we adopt $\Sigma _{\rm GMC}=85M_{\odot}{\rm pc}^{-2}$ and
$\sigma_{\rm g}=8$ km\,s$^{-1}$ for all galaxies following
\citet{2012ApJ...745...69K}.  From equations (\ref{eq:M_GMC2}) and
(\ref{eq:rho_GMC2}), we obtain the mass and density in the molecular
cloud regime using the value for $\Sigma _{\rm g}$ from
\citet{2012ApJ...745...69K}.

The obtained masses and densities for molecular clouds in the local
disk and starburst galaxies are presented in Figure \ref{fig:IC}.
Most galaxies are in the Toomre regime (and only 13 disk galaxies are
in the molecular cloud regime). The molecular
clouds typical for starburst galaxies are factors of 10 to 100 denser
than those following Larson's relation. Our massive and dense models
(m400k-d100, m400k-d10, and m1M-d100), which form young massive
clusters, are consistent with the molecular cloud observed in
starburst galaxies.  Starburst galaxies such as M83
\citep{2011MNRAS.417L...6B} and M51 \citep{2011ApJ...727...88C} are
indeed rich in dense massive clusters.  On the other hand, molecular
clouds typical in local disk galaxies do follow Larson's relation.
Our model m1M-d1, which forms classical open and leaky clusters
(associations), appears to be quite similar to these molecular clouds.

The typical molecular cloud in a disk galaxy, such as the Milky Way,
tends to form classical open clusters and associations, but these
clouds are insufficiently massive to form young massive star clusters.
This is consistent with the abundance of open star cluster and
associations in the Milky Way and with the lack of massive
star clusters.  According to our simulations, the formation of a
massive star cluster requires a massive ($\sim
10^5$--$10^6M_{\odot}$) and dense ($\sim$ 10--100$M_{\odot}$pc$^{-3}$)
molecular cloud.  Such a massive molecular cloud has, if virialized, a
velocity dispersion of $\sim 20{\rm km}\,{\rm s}^{-1}$.  Such a high
velocity dispersion (under compressive condition) could result from
the collision between two clouds
\citep{2009ApJ...696L.115F,2010ApJ...709..975O,2014ApJ...780...36F}.
Comparable high velocities are observed in the regions surrounding
young massive clusters, such as in the vicinity of NGC 3603
\citep{2014ApJ...780...36F} and Westerlund 2
\citep{2009ApJ...696L.115F,2010ApJ...709..975O}.  These clusters are
claimed to have been the result of cloud-cloud collisions
\citep{2009ApJ...696L.115F,2010ApJ...709..975O,2014ApJ...780...36F}.
These claims are supported by three-dimensional magneto-hydrodynamics
simulations, which also suggest that such cloud-cloud collisions
initiate the formation of massive cloud cores and potentially form
massive star clusters \citep{2013ApJ...774L..31I}.

Although our initial conditions of molecular clouds cover a relatively
wide range of mass and density, they are limited in our choice of
opting for homogeneous-density spheres. Recent numerical studies
indicate that molecular clouds with a concentrated density profile
such as a power-law, tend to form one high-mass star in the center
surrounded by many low-mass stars
\citep{2011MNRAS.413.2741G,2012MNRAS.420.3264G}. Such centrally
concentrated models then may more efficiently lead to the formation 
of massive clusters than our adopted homogeneous initial conditions.

\section{The mass and radius evolution of young star clusters}\label{Sect:MR_diagram}

Star clusters can be subdivided in several types, which represent
themselves clearly when presented in a mass-radius diagrams.  The
mass-radius distribution of star clusters changes with time. Here we
discuss the time evolution of the mass and radius of young clusters.

\subsection{Observations}

We start with summarizing the mass and radius evolution of observed
young star clusters.  These observations are presented in Figure
\ref{fig:M-R_diagram}, in particular for observed embedded clusters, 
classical open star clusters, young massive (starburst) clusters and
associations \citep{2003ARA&A..41...57L,2008A&A...487..557P,
  2009AJ....137.4777W,2003ApJ...593.1093L, 2006AJ....132.2296A,
  2009A&A...504..461F,2006ApJ...646.1215L,
  2008AJ....135..966F,2011MNRAS.414.3769B,1997AJ....113.1788H,
  1997MNRAS.286..538D,1991AJ....102.1108H,2009A&A...498L..37P,
  2010ARA&A..48..431P}.  For clarity we bin the clusters in age in
intervals of $t_{\rm age}=$ 1--5\,Myr, 5--20\,Myr, and 20--100 Myr.

\citet{2009A&A...498L..37P} and \citet{2010ARA&A..48..431P} list
several young massive clusters, but in many cases the listed radii
differ.  We adopt the half-mass radius given in
\citet{2010ARA&A..48..431P}, because the radius presented in
\citet{2009A&A...498L..37P} correspond to the core radius of the
clusters rather than the half-mass radius.  The former gives a more
direct comparison with our simulations.  In our analysis we try to
stay as much as possible to the same definition of cluster radius.
\citet{2008A&A...487..557P} present projected core and tidal radii by
fitting King models \citep{1966AJ.....71...64K}.  Because the density
profiles for the open clusters listed in \citet{2008A&A...487..557P}
are very shallow, we adopted their core radii, which for a King model
with $W_0=3$ is quite similar to the half-mass radius (the ratio of
the three-dimensional core radius to half-mass radius is 0.65 for a
King model with $W_0=3$).

Embedded clusters observed in the Milky Way galaxy reside almost
exclusively in the left panel of the mass-radius diagram ($t=1$--5 Myr
panel in Figure\,\ref{fig:M-R_diagram}) because they are young by
definition.  Embedded and classical open clusters populate the same
region (at the bottom left in the same panel). These clusters tend to
grow in size with age, which is a consequence of relaxation and out
gassing; embedded clusters observed in the Milky Way therefore appear
as ancestors of classical open clusters \citep{2015PASJ..tmp..163F}.
Associations populate the top right region of the left and middle
panels ($t=$1--5\,Myr and 5--20 Myr, respectively), and young massive
clusters are found to the right in the diagrams in
Figure\,\ref{fig:M-R_diagram}.  As was already suggested by
\citet{2009A&A...498L..37P}, young massive star clusters are well
separated in mass and radius from embedded and open clusters.  This
separation, however, diminishes for the older age group (20--100 Myr,
see the right panel of Figure \ref{fig:M-R_diagram}).

\subsection{Analytic model for the dynamical evolution of young star clusters}\label{Sect:Analytic}

The distribution and evolution of the observed star clusters in mass
and radius can be understood from out models of the dynamical
evolution for star clusters.

\begin{figure*}
\begin{center}
\includegraphics[width=54mm]{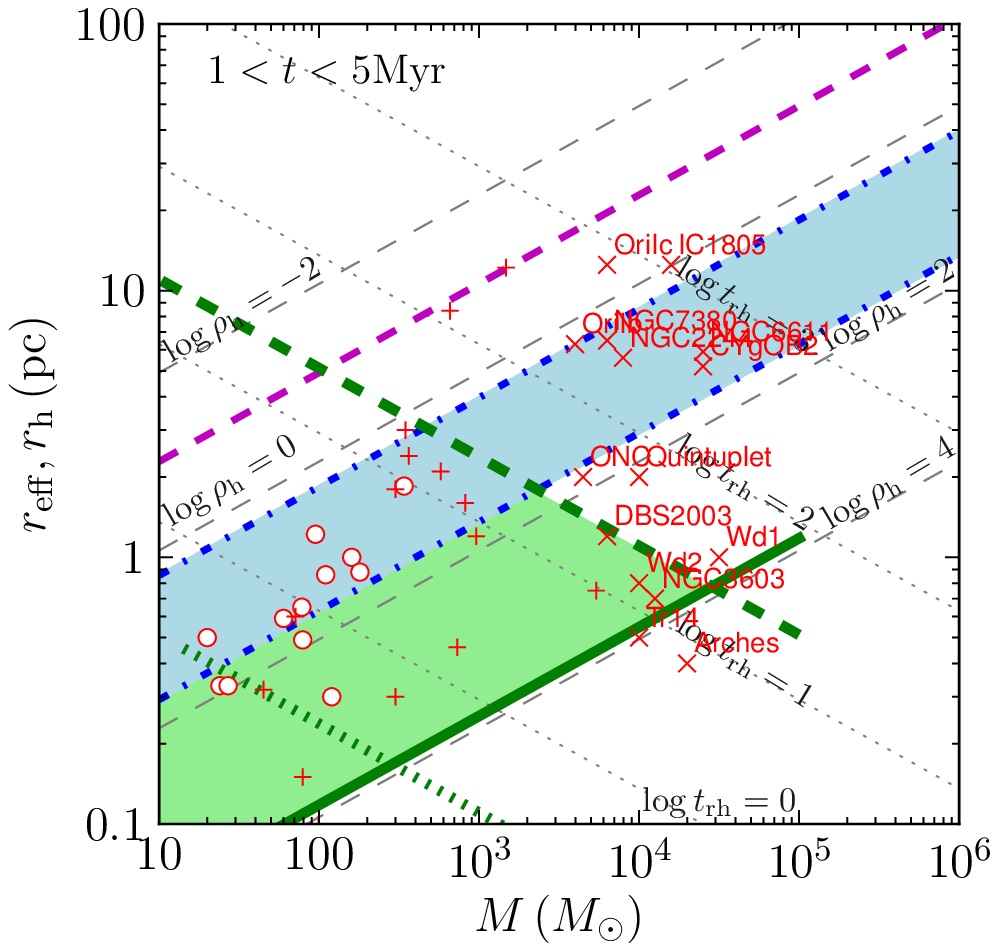}
\includegraphics[width=54mm]{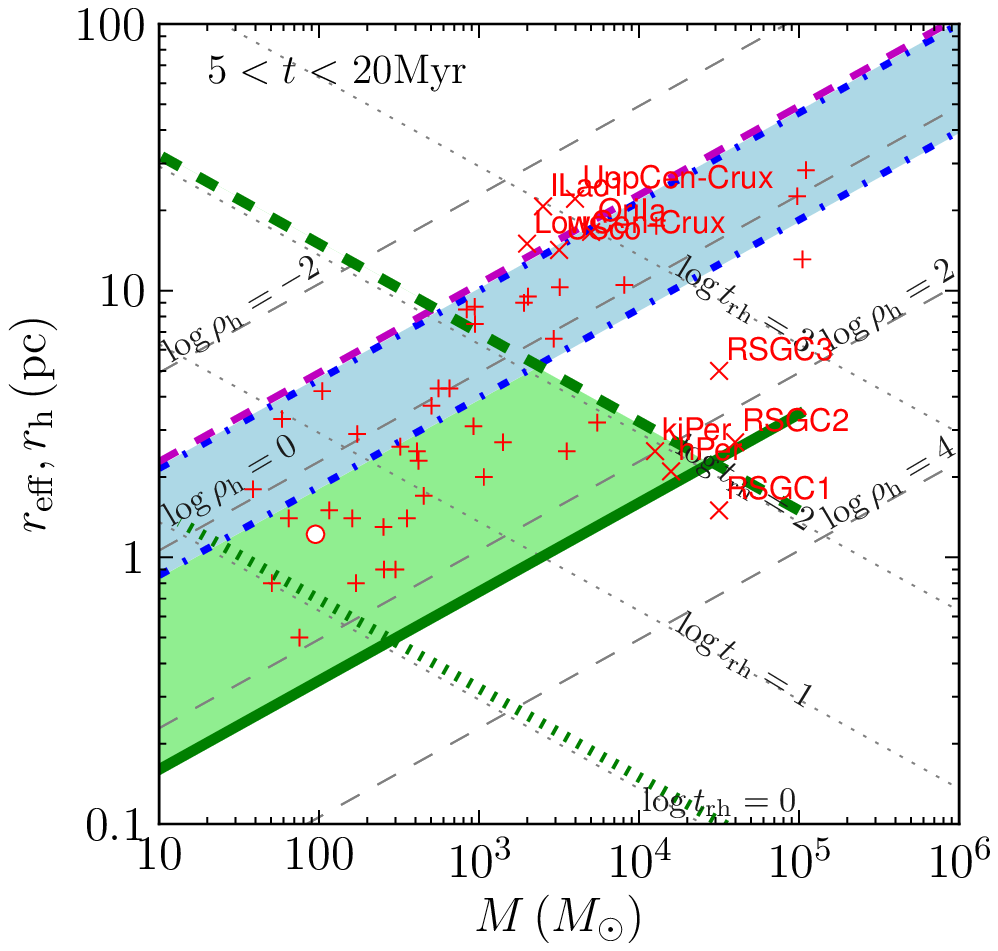}
\includegraphics[width=54mm]{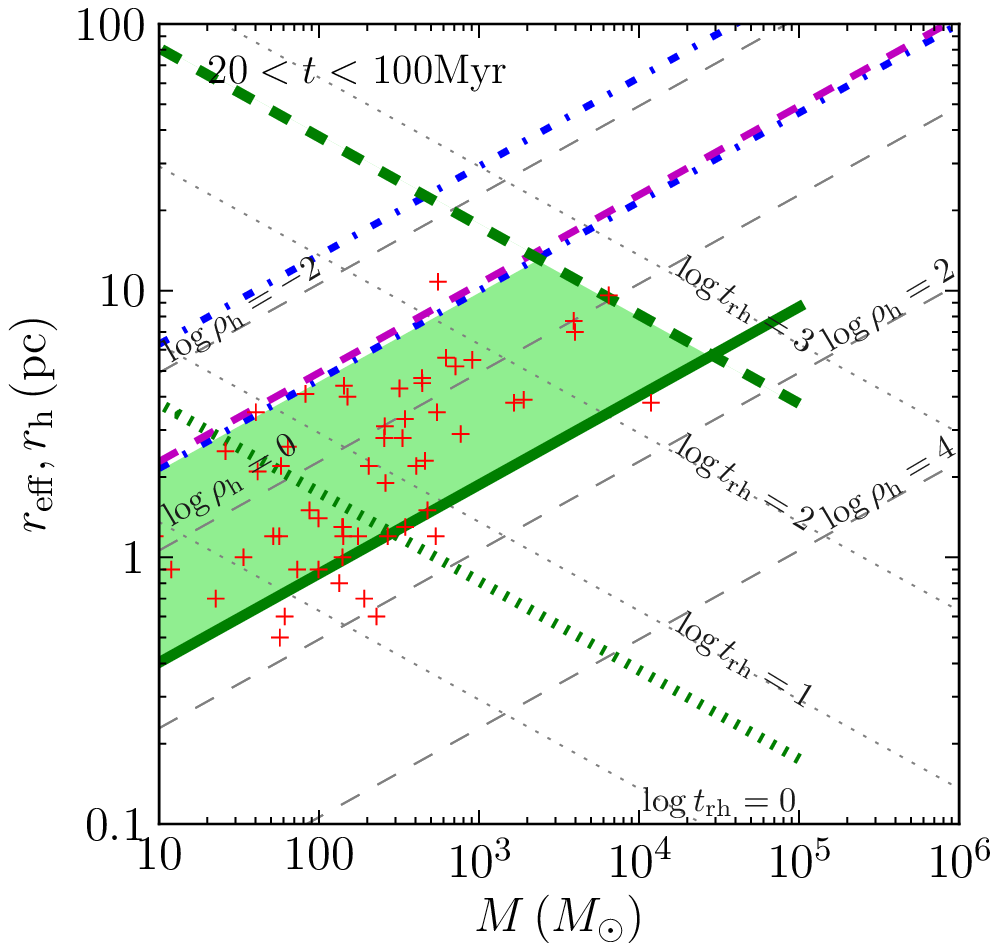}
\end{center}
\caption{Mass-radius diagrams of observed young star clusters for
  $t_{\rm age}=$1--5, 5--20, and 20--100 Myr from left to right. The data is
  from \citet{2003ARA&A..41...57L} for embedded clusters (red
  circles),
\citet{2008A&A...487..557P,2009AJ....137.4777W,2003ApJ...593.1093L,2006AJ....132.2296A,
2009A&A...504..461F,2006ApJ...646.1215L,2008AJ....135..966F,2011MNRAS.414.3769B,1997AJ....113.1788H,1997MNRAS.286..538D,1991AJ....102.1108H}
  for open clusters(red pluses), and \citet{2010ARA&A..48..431P} for
  young massive clusters and leaky clusters (red crosses). The data
  for the leaky clusters are overlapped with
  \citet{2009A&A...498L..37P}. The clusters listed in
  \citet{2010ARA&A..48..431P} are shown with the names.
\label{fig:M-R_diagram}}
\end{figure*}

The lower limit of the cluster density can be understood by considering 
the background density in the field.
The magenta dashed line in the diagram indicate $\rho = 0.1M_{\odot}{\rm
  pc}^{-3}$, which is an order of magnitude higher than the mean
density of the field stars in the solar neighborhood
\citep{2000MNRAS.313..209H}.  We adopt $\rho = 0.1M_{\odot}{\rm
  pc}^{-3}$ as a lower limit for the cluster density 
(magenta dashed line in Figure \ref{fig:M-R_diagram}).  Star clusters
with a density similar to or lower than the mean stellar density would
therefore not be recognizable as clusters. And indeed, only a few of
the most massive clusters reside above this curve, and those have a
relatively high concentration. As a consequence, their core densities
exceed the local density considerably, which helps to identify them
as clusters in observational campaigns.

The blue dash-dotted lines in Figure \ref{fig:M-R_diagram} indicate
the mass-radius relation for which the dynamical time scale (see
equation (\ref{eq:t_dyn})) is equal to the age of the cluster.  Each
panel contains two lines, one for the minimum and one for the maximum
age of the clusters shown in the panels.  
The regions between these lines is shaded blue.
Clusters between or below the blue lines will be recognizable as a bound
systems unless the lines exceed the magenta dashed line.  
\citet{2010ARA&A..48..431P} and \citet{2011MNRAS.410L...6G}
argued that the ratio between cluster age and the dynamical time
provides a good indicator for separating the bound from the unbound
systems: They adopt as a criterion $t_{\rm age}/t_{\rm dyn} \gtrsim 3$
to make this distinction. Using this criterion they categorized the
leaky clusters in \citet{2009A&A...498L..37P} as associations. The
blue region in Figure \ref{fig:M-R_diagram} moves upward with time,
together with the observed clusters.  At $t>20$ Myr (the right panel
in Figure \ref{fig:M-R_diagram}) the blue lines are located above the
magenta line, indicating that these clusters have a density too low to
be recognized as a cluster. 

The evolution of dense star clusters is quite different from those of
open clusters or associations.  Dense star cluster evolution can
roughly be divided in two phases; before core collapse and after core
collapse.  In the former phase the core radius of the star clusters
shrinks, and as a consequence, its core density increases
\citep{1965AnAp...28...62H,1968MNRAS.138..495L}.  From the moment the
first hard binaries form in the cluster core
\citep{1971ApJ...164..399S,1974A&A....35..237A}, they act as an energy
sources \citep{1975MNRAS.173..729H,1983ApJ...272L..29H} causing the
core to re-expand.  From this moment on, the core- and half-mass
radius of clusters increases.  These processes proceed on the
half-mass relaxation time:
\begin{eqnarray}
  t_{\rm rh} = \frac{0.065 \sigma^3}{G^2 \langle m\rangle \rho \ln \Lambda}.
\label{eq:trlx}
\end{eqnarray}
Here $\sigma$ and $\rho$ are the velocity dispersion and density of
the cluster, respectively, and $\ln \Lambda$ is the Coulomb logarithm
\citep{1987degc.book.....S}.  We rewrite equation\,(\ref{eq:trlx}) to
include some common dimensions in equation (\ref{eq:t_rh}).

\citet{2011MNRAS.413.2509G} modeled the post-collapse evolution of the
half-mass radius and the density of star clusters due to the energy
flux from the core following the description of
\citet{1965AnAp...28...62H}. We attempt to understand the dynamical
evolution of young star clusters using their description.  We ignore
the pre-collapse phase and consider only the evolution in the
post-collapse (expansion) phase, because the pre-collapse phase is
much shorter than post-collapse. The core-collapse time, which is
the time for the pre-collapse phase, scales with the relaxation time
(see equation (\ref{eq:t_rh}) or (\ref{eq:trlx})).  
This time scale depends on the stellar mass
function, and for clusters with a realistic mass function the core
collapse time is generally shorter than one relaxation
\citep{2002ApJ...576..899P,2004ApJ...604..632G,2014MNRAS.439.1003F}.
Since most of young open clusters in our observed sample have a
relaxation time $\lesssim 10$\,Myr (see the left panel of Figure
\ref{fig:M-R_diagram}), they probably reach core collapse well within
a few Myr.  We also ignore the effect of the Galactic tidal field,
because the time scale we treat here is short ($<100$ Myr) compared to
the time scale for the tidal disruption ($\sim 1$Gyr)
\citep{2007MNRAS.376..809G}.

The time of the half-mass radius of clusters due to binary heating in
the core is given by equation (B7) in \citet{2011MNRAS.413.2509G}:
\begin{eqnarray}
   r_{\rm h}\simeq\left( \frac{3G}{4\pi N}\right)^{1/3} (125\zeta t)^{2/3}.
\label{eq:Gieles_model}
\end{eqnarray}
Here $N$ is the initial number of stars in the cluster. In equation
(B7) in \citet{2011MNRAS.413.2509G}, the cluster mass is assumed that
$M=\langle m \rangle N$, where $\langle m \rangle$ is the mean
stellar mass. They adopted a scaled mass $\langle m \rangle=0.5$ and 
as a result their equation (B7) is slightly different from our equation.
If we assume that $\langle m \rangle = 0.5\,M_{\odot}$, we can write
this equation as
\begin{eqnarray}
r_{\rm h} \simeq 2.0\, \zeta ^{2/3}\left(\frac{M}{M_{\odot}}\right)^{-1/3}
\left( \frac{t_{\rm age}}{{\rm Myr}} \right)^{2/3}  \,{\rm pc}.
\label{eq:Gieles_model_2}
\end{eqnarray}
Here we adopt $t=t_{\rm age}$ because we ignore the pre-collapse phase.
The the expansion-rate coefficient, $\zeta$, depends on the ratio of
the maximum to the minimum mass in the stellar mass function, $\mu
\equiv m_{\rm max}/m_{\rm min}$.  In \citet{2011MNRAS.413.2509G} we
use $\zeta\simeq0.2$, which corresponds to $\mu \simeq 10$, and which
is appropriate for globular clusters.  Young clusters however, should
have a larger value of $\mu$ because of the presence of massive stars.
For some of these clusters $m_{\rm max}/m_{\rm min} \simeq
100\,M_{\odot}/0.01\,M_{\odot} \simeq 10^4$. Following
\citet{2011MNRAS.413.2509G} and assuming $\zeta \propto \mu^{1/2}$, we
obtain $\zeta \simeq 20$ for $\mu \simeq 10^4$.  Equation
(\ref{eq:Gieles_model_2}) with $\zeta=20$ for $t=t_{\rm age}=2$, 10,
40\,Myr is shown as green dashed lines in Figure
\ref{fig:M-R_diagram}.

The majority of the observed clusters are located below this
evolutionary line rather than straddling the line, which indicates
that they have $\zeta<20$. Green dotted lines in each panel of Figure
\ref{fig:M-R_diagram} shows equation (\ref{eq:Gieles_model_2}) with
$\zeta=0.2$. Most of observed embedded and classical open clusters are
located between the dotted green (for $\zeta=0.2$) and the dashed
green ($\zeta = 20$) lines.  This may be caused by the large dispersion
in $\zeta$, as we discussed here, or because the pre-collapse time is
not taken into account in our analysis. By ignoring the pre-collapse
time we reduce the evolution time of a star cluster compared to the
expectation.

The descriptions of \citet{2011MNRAS.413.2509G} (equations
(\ref{eq:Gieles_model}) and (\ref{eq:Gieles_model_2})) give infinite
density at $t=0$ Myr, which hardly seems realistic for actual young
star clusters.  Instead, we adopt equation (B4) of
\citet{2011MNRAS.413.2509G}:
\begin{eqnarray}
\rho_{\rm h}\simeq \frac{1}{G}\left( \frac{N}{250\zeta t} \right)^2,
\label{eq:Gieles_density}
\end{eqnarray}
which gives the half-mass density as a function of time. 
We also adopt $\langle m \rangle = 0.5\,M_{\odot}$.  
We assume a (maximum) half-mass density
of $10^4M_{\odot}{\rm pc}^{-3}$ at $t_{\rm age}=0.1$ Myr irrespective
of the cluster mass; we obtain $\rho_{\rm h}(M_{\odot}{\rm
  pc}^{-3})=100(t/{\rm Myr})^{-2}$.  Since $\rho_{\rm h}=3M/(8\pi
r_{\rm h}^3)$, the relation can be written as
\begin{eqnarray}
  r_{\rm h} = 0.0158\left(\frac{M}{M_{\odot}}\right)^{1/3}
  \left(\frac{t_{\rm age}}{{\rm Myr}}\right)^{2/3} \, {\rm (pc)}.
\label{eq:model}
\end{eqnarray}

Equation\,(\ref{eq:model}) is presented in Figure \ref{fig:M-R_diagram}
as the solid green line.  One naively expects that clusters with an
initial density smaller than $10^4M_{\odot}{\rm pc}^{-3}$ populate the
area above this line, which is consistent with the observations. From
a theoretical perspective we argue that star clusters are expected to
reside in the green and blue regions in Figure \ref{fig:M-R_diagram},
which for the majority of observed clusters appear to be the case.

From these results, the regions in which clusters are expected
to exist on mass-radius diagrams are shown by green and blue shades
in Figure \ref{fig:M-R_diagram}, and observed distribution of 
star clusters matches them. Furthermore, our analytic models 
suggest two distinct populations of massive ($\sim 10^4M_{\odot}$) 
clusters, which are called as starburst and leaky clusters by
\citet{2009A&A...498L..37P}. We argue that these two populations
naturally appear if we consider the formation and the dynamical 
evolution process of star clusters.

\subsection{Time evolution of cluster radius: Leaky and starburst clusters}

Young star clusters with $M\sim10^4M_{\odot}$ are divided into two groups,
as can be seen in Figure \ref{fig:M-R_diagram}.
\citet{2009A&A...498L..37P} named them starburst (young massive)
clusters and leaky clusters (following \citet{2010ARA&A..48..431P} we
identify the latter category as associations).
\citet{2009A&A...498L..37P} and \citet{2011A&A...536A..90P} showed
that both families of clusters expand with time, but at a different
rate: $r/{\rm pc}=0.16(t_{\rm age}/{\rm Myr})$ for the starburst
clusters and $r/{\rm pc}=3.5(t_{\rm age}/{\rm Myr})^{2/3}$ for the
associations.  In this section, we discuss the origin of these different
evolutionary tracks.

\begin{figure*}
\begin{center}
\includegraphics[width=80mm]{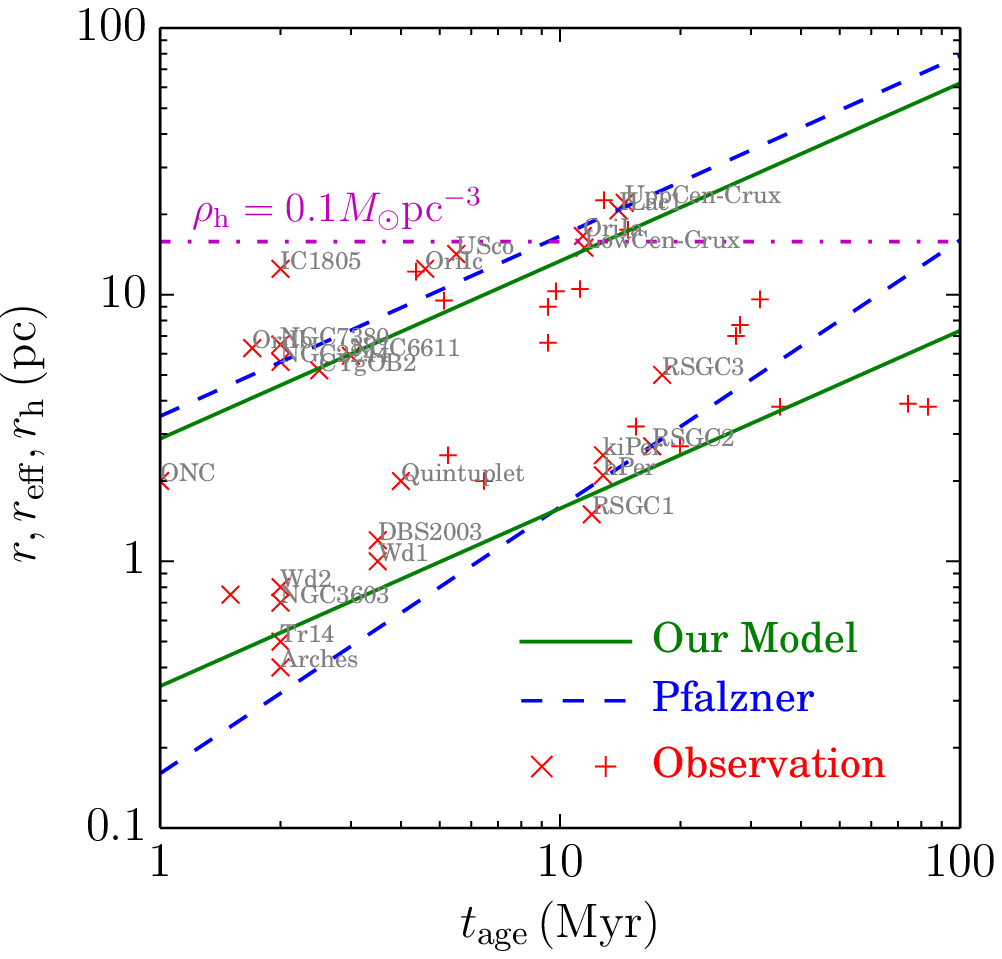}
\end{center}
\caption{Cluster radius as a function of time for clusters with a mass
  of $10^3$--$10^5M_{\odot}$. Red plus-signs are from
  \citet{2010ARA&A..48..431P} and red pluses from the others (see the
  caption of Figure \ref{fig:M-R_diagram}). Blue dashed lines are the
  relation given in \citet{2011A&A...536A..90P} and
  \citet{2009A&A...498L..37P}: $r=3.5t_{\rm age}^{2/3}$ and $r=0.16t_{\rm age}$ for top
  and bottom, respectively.  Green lines are cluster radii as a
  function of time obtained from our model: $r_{\rm h}=2.7t_{\rm age}^{2/3}$ and
  $r_{\rm h}=0.34t_{\rm age}^{2/3}$ for top and bottom, respectively.  Note that
  for starburst clusters, we plot the half-mass radii given in
  \citet{2010ARA&A..48..431P} instead of ``size'' in
  \citet{2009A&A...498L..37P}.
\label{fig:t-r}}
\end{figure*}

In Figure \ref{fig:t-r} we present the age and radius of observed in
young star clusters with a mass of $10^3<M<10^5M_{\odot}$.  The time
evolution for cluster radii, plotted as the solid green lines, are
obtained from the analytic models discussed in \S\,\ref{Sect:Analytic}.

Associations are about one dynamical time scale old, and therefore we
can hardly confirm weather they are bound or not.  If we consider them
to be one dynamical timescale old, i.e., $t_{\rm age} \simeq t_{\rm
  dyn}$, equation (\ref{eq:t_dyn}) gives $r_{\rm h}/{\rm
  pc}=2.7(t_{\rm age}/{\rm pc})^{2/3}$.  We present this relation in
Figure \ref{fig:t-r} as the top green line.  The model is consistent
with the observed clusters, and the power-law index of our model is
consistent with that of \citet{2011A&A...536A..90P}.

For starburst clusters, we adopt the results based on
\citet{2011MNRAS.413.2509G}.  By adopting $M=10^4M_{\odot}$ in
equation (\ref{eq:model}), we obtain $r_{\rm h}/{\rm pc}=0.34(t_{\rm
  age}/{\rm Myr})^{2/3}$.  We present this relation in Figure
\ref{fig:t-r} as the bottom green line. In part due to the large
scatter, this relation is also consistent with the observed radius
evolution of starburst clusters.

The magenta dash-dotted line in Figure \ref{fig:t-r} gives the
relation $\rho_{\rm h}=0.1M_{\odot}{\rm pc}^{-3}$.  This is an order
of magnitude higher than the field density at solar neighborhood
($0.01M_{\odot}{\rm pc}^{-3}$) \citep{2000MNRAS.313..209H} and we
assume this to be a minimum to the (observable) cluster density.
This predict that associations will not survive more than $\sim 20$
Myr, and indeed no such a cluster has been observed.

\section{Summary}

We performed a series of simulations of star forming regions.  Our
calculations start with hydrodynamical simulations of turbulent
molecular clouds.  These simulations are continued for about one
initial free-fall time scale, after which we replace gas particles
with stars adopting a local star formation efficiency
\citep{2012ApJ...745...69K}. The stellar mass are selected randomly
from the adopted initial mass function, and the stars receive the
position and velocity of the gas particles they replace.  We
subsequently removed all residual gas and continue the evolution of
the young emerging star cluster by means of $N$-body simulations with
stellar evolution.

The types of star clusters that formed in our simulations depend on
the initial conditions (mass and density) of the molecular cloud.  The
clouds with initial conditions typical for those observed in the Milky
Way ($10^4M_{\odot}$ and 100--1000 cm$^{-3}$) lead to classical open
clusters.  More massive clouds ($10^5$--$10^6M_{\odot}$) with the same
density evolve into dense massive clusters. These massive molecular
clouds are common in starburst galaxies, but are very rare in local
disk galaxies such as the Milky Way.  This result is consistent with
observations that young massive clusters are common in starburst
galaxies, but only several has been found in the Milky Way.  We argue
that such massive clouds must be able to form in the Milky Way Galaxy,
even though they are probably rare.

Dense massive clusters in our simulation form from molecular 
clouds with a mass of $10^6M_{\odot}$ and a density of 
$\sim 1000$ cm$^{-3}$ ($100M_{\odot}{\rm  pc}^{-3}$) leading to a 
velocity dispersion of $\sim 20$ km\,s$^{-1}$.  This is consistent 
with the relative velocity of
molecular clouds observed near young massive clusters in the Milky Way
such as near NGC 3603 \citep{2014ApJ...780...36F} and Westerlund 2
\citep{2009ApJ...696L.115F,2010ApJ...709..975O}.  We argue that
massive clusters in the Milky Way can therefore not form from
individual clouds, but their formation may have been initiated in
cloud-cloud collisions
\citep{2009ApJ...696L.115F,2010ApJ...709..975O,2014ApJ...780...36F}.

Molecular clouds with a mass of $\sim10^6M_{\odot}$ and a low density
of $\sim10$\,cm$^{-3}$ ($\sim 1M_{\odot}{\rm pc}^{-3}$), which follow
Larson's relation, tend to form associations (``leaky clusters'' in the
terminology of \citet{2009A&A...498L..37P}).  These relatively low
density and massive molecular clouds form a number of small
clumps. They might be detected as embedded or classical open clusters
when they are young, but they evolve to less dense clusters due to the
gas expulsion and relaxation. After several Myr, these systems lose
their clumpiness and become recognizable as associations.

In our simulations we assumed that stars form instantaneously upon the
expulsion of the residual gas (after an initial free-fall time of the
molecular cloud).  Our prescription for star formation is simple
compared to reality, in which star formation triggers the expulsion of
the residual gas by means of feedback processes. Regardless the
simplicity of our approach, we are still able to make a distinction
between the formation of associations, open clusters, and massive star
clusters.

The young stellar system, Sco OB2, is an assembly of associations of
slightly different ages, USco, Upper Cen-Lups, and Lower Cen-Crux. A
stellar system similar to Sco OB2 naturally originates in our
simulations of relatively massive and low-density molecular clouds,
although the age spread cannot be reproduced with our method.  The
relation that less dense clusters have wider age spreads of stars is
observationally and theoretically suggested
\citep{2014ApJ...791..132P}.

In addition, we compared our simulations with theoretical models for
cluster expansion due to the dynamical evolution
\citep{2011MNRAS.413.2509G}.  These models satisfactorily explain the
evolution in radius of simulated clusters as well as the observed
clusters.

We also found that the distribution of clusters on the mass-radius
diagram is also limited by the density with which the dynamical time
scale is equal to the cluster age. This implies that if the cluster
age is much shorter than the dynamical time; such clusters cannot be
recognized as (bound) systems \citep{2011MNRAS.410L...6G}.  After
$\simeq 20$ Myr the density of these associations drops below the
background density and dissolve.

The gap of the radius distribution for associations and young massive
clusters suggested by \citet{2009A&A...498L..37P} is consistent with
our simulation results. While young massive clusters evolve following
the cluster expansion model, leaky clusters have $t_{\rm age}\sim
t_{\rm dyn}$. With our models, the evolution of radius for observed
leaky and young massive clusters are described by $r_{\rm h}/{\rm
  pc}=2.7(t_{\rm age}/{\rm pc})^{2/3}$ and $r_{\rm h}/{\rm
  pc}=0.34(t_{\rm age}/{\rm Myr})^{2/3}$, respectively.  These are
also consistent with observations.  \citet{2014ApJ...794..147P}
claimed that star formation continues in embedded clusters and that
after the gas expulsion they expand and become associations. Our
models however indicate that clumpy star forming regions are observed
as a conglomerate of embedded clusters, but at a later time these
systems lose their clumpiness due to the expulsion of the residual gas
and two-body relaxation.  Because our coverage of parameter space
remains limited and much is still to be uncovered, we hope to explore
a much wider range of initial conditions of molecular clouds
(different masses, radii, and density distributions) and other
assumption for star formation (different epochs for star formation and
gradual gas removal rather than instantaneous gas expulsion).

Our results suggest that the difference in the parental molecular
clouds results in the formation of various types of star clusters if
we assume the same star formation process and that the cluster
formation process does not depend on the condition of the galaxy,
either normal disk or starburst.

\acknowledgments

We thank the anonymous referee for useful comments.
This work was supported by JSPS KAKENHI Grant Number 26800108 and NAOJ
Fellowship, the Netherlands Research Council NWO (grants
\#643.200.503, \#639.073.803 and \#614.061.608), and the Netherlands
Research School for Astronomy (NOVA).  Numerical computations were
partially carried out on Cray XC30 at the Center for Computational
Astrophysics (CfCA) of the National Astronomical Observatory of Japan
and Little Green Machine at Leiden Observatory.

\bibliographystyle{apj}
\bibliography{reference}

\end{document}